\documentclass[12pt]{article}
\usepackage{epsfig}
\usepackage{color,graphicx}
\usepackage{cite}
\usepackage{amssymb,amsmath}

\newcommand{\be}{\begin{equation}}
\newcommand{\ee}{\end{equation}}
\newcommand{\bea}{\begin{eqnarray}}
\newcommand{\eea}{\end{eqnarray}}

\newcommand{\gsim}{\gtrsim}

\begin{document}

\begin{flushright}
HIP-2017-11/TH\\
\end{flushright}

\begin{center}  

\vskip 2cm 

\centerline{\Large {\bf Striped anyonic fluids}}
\vskip 1cm

\renewcommand{\thefootnote}{\fnsymbol{footnote}}

\centerline{
Niko Jokela,${}^{1,2}$\footnote{niko.jokela@helsinki.fi} 
Gilad Lifschytz,${}^{3,4}$\footnote{giladl@research.haifa.ac.il} 
and Matthew Lippert${}^5$\footnote{Matthew.Lippert@liu.edu}}

\vskip 0.5 cmÍ

\vspace{4mm}
${}^1${\small \sl Department of Physics} and ${}^2${\small \sl Helsinki Institute of Physics} \\
{\small \sl P.O.Box 64} \\
{\small \sl FIN-00014 University of Helsinki, Finland} 

\vspace{2mm}
\vskip 0.2cm
${}^3${\small \sl Department of Mathematics, and} \\
${}^4${\small \sl Haifa Research Center for Theoretical Physics and Astrophysics} \\
{\small \sl University of Haifa, Haifa 31905, Israel}

\vspace{2mm}
\vskip 0.2cm
${}^5${\small \sl Department of Physics} \\
{\small \sl Long Island University Brooklyn} \\
{\small \sl Brooklyn, NY 11201, USA} 

\end{center}

\vskip 0.3 cm

\setcounter{footnote}{0}
\renewcommand{\thefootnote}{\arabic{footnote}}

\begin{abstract}
\noindent 
The choice of statistics for a quantum particle is almost always a discrete one: either bosonic or fermionic.  Anyons are the exceptional case for which the statistics can take a range of intermediate values. Holography provides an opportunity to address the question of how the behavior of interacting anyons depends on the choice of statistics. In this paper, we analyze the spectrum of a strongly coupled, gapless fluid of anyons described holographically by the D3-D7' model with alternative boundary conditions. 
We investigate how these alternative boundary conditions 
impact the instability of the gapless homogeneous phase toward the formation of spatial order. 
In addition, we also show that for a particular, limiting choice of the alternative boundary conditions, this holographic system can be interpreted as describing strongly coupled $(2+1)$-dimensional QED.  In this case, the instability leads to a spontaneous, spatially modulated magnetic field.

\end{abstract}

\newpage

\section{Introduction}

To those of us living in a higher numbers of dimensions, the physics of two spatial dimensions can be unusual and surprising.  Quantum particles, in addition to being bosons or fermions, can also be anyons.  Rather than being restricted to integer or half integer spins, anyons can have any spin in between.  Exchanging two identical anyons shifts the wave function by a phase which is not just zero or $\pi$ but can take any value from zero to $2\pi$.  Rather than just a theoretical curiosity, anyons are found as quasiparticles in fractional quantum Hall states \cite{Laughlin:1983fy} and have even been observed experimentally \cite{anyonexperiments}.  Furthermore, a gas of anyons forms a unconventional superfluid \cite{Fetter:1989fu,Lee:1989fw,Chen:1989xs,Lykken:1989ym}, which led to a brief role as a possible explanation for high $T_c$ superconductivity \cite{anyonshighTc}.

However, forty years after their discovery, anyons remain quite poorly understood \cite{openproblems}.  Even in such a basic system as the non-interacting anyon gas, many basic questions remain unanswered.  For example, what is the ground state for more than two anyons?  Does an anyon gas ever Bose condense?  Is there a virial expansion to describe the low-density equation of state?  What is the low-energy effective hydrodynamic description of an anyon gas at high density and low temperature?

Gauge/gravity duality is well positioned to answer many of these open questions.  Some intractable quantum many-body problems can be rendered nearly trivial in the holographic dual description.  For instance, finite density fermion systems can be analyzed holographically without any trace of the fermion sign problem, which severely impedes numerical lattice computations.  Anyon fluids may prove to be another such powerful application of holography.  

Fractional statistics can be obtained in a $(2+1)$-dimensional CFT with a conserved $U(1)$ by an $SL(2,{\mathbb{Z}})$ mapping which mixes the $U(1)$ current and the background vector field \cite{Witten:2003ya, BurgessandDolan}.  This procedure is easily implemented holographically by considering alternative quantization of the bulk gauge field \cite{Jokela:2013hta, Brattan:2013wya,Ihl:2016sop}.  As usual, a semiclassical gravity dual describes a strongly coupled anyon fluid.  This seems to indicate that for anyons, perhaps, it is in the strong-coupling rather than in the weak-coupling limit where the physics is simplest.   In fact, recently the last two of the above questions have been partially addressed in \cite{Ihl:2016sop}.

Alternative quantization can instead be interpreted as a way to include dynamical gauge fields in the boundary theory.  Rather than changing the statistics of the particles, the transformation can be viewed as changing which combination of the gauge field and charged currents is held fixed and which combination is allowed to fluctuate.

Holographic anyons have been recently studied in a variety of probe brane contexts \cite{Jokela:2013hta, Brattan:2013wya, Jokela:2014wsa, Brattan:2014moa, Jokela:2015aha, Itsios:2015kja, Itsios:2016ffv,Jokela:2016nsv}.  
A particularly interesting example is the D3-D7' system, originally constructed as a model for the fractional quantum Hall effect \cite{Bergman:2010gm}, which was generalized in \cite{Jokela:2013hta} to describe a system of anyons.  The gapped quantum Hall state can be transformed by an appropriate $SL(2,{\mathbb{Z}})$ mapping into an anyonic superfluid, whose properties and instabilities were described in \cite{Jokela:2014wsa}.

In this paper, we focus on another feature of the D3-D7' model: the spatially modulated instability.  In many physical systems, the ground state at finite charge density features spontaneously broken spatial symmetries.   Often, the instability of the homogeneous state is enhanced for large charge and is suppressed by a magnetic field. 

The gapless phase of the D3-D7' model is a typical example; at sufficiently high density, the homogeneous state is unstable to fluctuations at nonzero momentum \cite{Bergman:2011rf, Jokela:2012vn}.  The end point of this instability is a striped spin, charge, and current density wave, with a spatial modulation closely matching the wavelength of the unstable modes \cite{Jokela:2014dba}.  At zero magnetic field, the phase transition between homogeneous and striped states is second order.  When a magnetic field is applied, this transition becomes first order and the critical charge density increases.

Here, we examine the behavior of this modulated instability under an $SL(2,{\mathbb{Z}})$ transformation mapping the fluid of fermions into one composed of anyons.  Recently, it was argued that a uniform fluid of free anyons at low density is unstable to becoming inhomogeneous \cite{Hudak}.  We show here that this strongly coupled anyonic fluid has an instability towards a striped phase and that
the $SL(2,{\mathbb{Z}})$ transformation 
has a notable effect both on where in parameter space the instability occurs and on the nature of the modulated ground state. Compared with the original fermion fluid, 
the wavelength of the unstable fluctuation is longer.
Furthermore, the instability for the anyons is typically driven by the hydrodynamical diffusion mode.

 A particularly interesting case is the S transformation of the original CFT.  Rather than interpreting this mapping as altering the statistics of the particles, one can instead view it as changing the background gauge field into a dynamical one.  The resulting theory is essentially $(2+1)$-dimensional strongly coupled QED with a large number of flavors.  At weak coupling, $(2+1)$-dimensional QED at nonzero density is thought to undergo spontaneous magnetization \cite{Zeitlin:1996nr, Zeitlin:1996ys, Kanemura:1996yd, Eliashvili:2000yd}.  In the strongly coupled case, there is an instability above a critical density to form a spontaneously modulated magnetic field.  We are able to numerically solve for this nonlinear, striped ground state, in a calculation similar to \cite{Jokela:2014dba}.

We organize the discussion in this paper as follows.  We first review in Sec.~\ref{sec:D3D7} the construction of the D3-D7' system, the alternative boundary conditions which yield an anyonic fluid, and the set-up of the fluctuation analysis.  Then, in Sec.~\ref{sec:anyonicstripes}, we study the quasinormal mode spectrum, in particular the onset of the modulated instability.  We focus in Sec.~\ref{sec:2+1QED} on the S transformation of the original system, interpreting it as the holographic dual of strongly coupled, $(2+1)$-dimensional QED.  Finally, we discuss the results, remaining open questions, and outlook for the future in Sec.~\ref{sec:discussion}.


\section{D3-D7' system}
\label{sec:D3D7}

We begin by reviewing the D3-D7' system and its generalization to describe a strongly coupled anyon fluid.  The original model was constructed by embedding a probe D7-brane in a D3-brane background such that the intersection is a $(2+1)$-dimensional defect which breaks supersymmetry and whose low-energy excitations are purely fermionic.  This model has a well studied phenomenology, with both a Minkowski embedding, yielding a gapped, quantum Hall phase, and a black hole embedding, dual to a gapless, conducting phase \cite{Bergman:2010gm, Jokela:2010nu, Bergman:2011rf, Jokela:2012vn, Jokela:2014dba, Davis:2011gi, Omid:2012vy}. Other closely related models include \cite{Alanen:2009cn,Jokela:2011eb,Jokela:2011sw,Jokela:2012se,Hutchinson:2014lda,Bea:2014yda,Mezzalira:2015vzn,Kristjansen:2016rsc}.

By considering alternative quantization for the D7-brane gauge field, the fermions become anyons.  The anyonization of the Minkowski embedding was explored in \cite{Jokela:2013hta, Jokela:2014wsa}; this case does not suffer from a modulated instability and, for a specific choice of statistics, describes an anyonic superfluid.  In this paper, we will instead focus on the instability of the anyonized black hole embedding.

\subsection{Background}

The near-horizon background of a stack of $N_c$ non-extremal D3-branes is
\begin{eqnarray}
\label{D3metric}
 L^{-2} ds_{10}^2 &=& r^2 \left(-h(r)dt^{2}+dx^2+dy^2+dz^2\right)+
 r^{-2} \left(\frac{dr^2}{h(r)}+r^2 d\Omega_5^2\right) \\
\label{RR_5-form}
F_5 &=& 4L^4\left(r^3 dt\wedge dx\wedge dy\wedge dz\wedge dr 
+  d\Omega_5 \right) \,,
\end{eqnarray}
where $h(r)=1-r_T^4/r^4$ and $L^2=\sqrt{4\pi g_{s} N_c}\, \alpha'$. 
For convenience, we work in dimensionless coordinates, e.g., $r=r_{phys}/L$.
This background is dual to ${\cal N}=4$ SYM theory at a temperature $T = r_T/(\pi L)$. 

We parameterize the internal five-sphere as an $S^2\times S^2$ fibered over an interval:
\be
 d\Omega_5^2 = d\psi^2 + 
 \cos^2\psi \left(d\theta^2 + \sin^2\theta d\phi^2 \right) + \sin^2\psi \left(d\alpha^2 + \sin^2\alpha d\beta^2 \right)  \,,
\ee
where $\psi \in [0,\pi/2]$, $\theta, \alpha \in [0, \pi]$, and $\phi, \beta \in [0, 2\pi]$.
As $\psi$ varies, the sizes of the two $S^2$'s change. At $\psi=0$ one of the $S^2$'s shrinks to zero
size, and at $\psi=\pi/2$, the other $S^2$ shrinks. The $S^2\times S^2$ at $\psi=\pi/4$ is the ``equator" of the $S^5$.

\subsection{D7-brane probe}

The probe D7-brane extends in the $t,x,y$, and $r$ directions and wraps the two two-spheres; it represents a $(2+1)$-dimensional defect in the $(3+1)$-dimensional spacetime directions.  We initially consider a homogeneous embedding, depending only on $r$ and characterized by two functions, $\psi(r)$ and $z(r)$.  We also include a homogeneous nonzero charge density and background magnetic field by turning on the following D7-brane worldvolume gauge fields:
\bea
F_{xy} &=&  \frac{L^2}{2\pi \alpha' }  b \\ 
F_{rt} &=&  \frac{L^2}{2\pi \alpha' }  {a_t}' \ .
\eea
The physical magnetic field is given by $B  =  \frac{1}{2\pi \alpha' }  b$. 

This embedding completely breaks supersymmetry and is unstable to the D7-brane slipping off the $S^5$.  We cure this instability by turning on $f_1$ and $f_2$ units of magnetic flux through the two two-spheres:
\bea
F_{\theta\phi} &=& \frac{L^2}{2\pi \alpha' } \frac{f_1}{2} \sin\theta \\
F_{\alpha\beta} &=&  \frac{L^2}{2\pi \alpha' }  \frac{f_2}{2} \sin\alpha \ .
\eea
For $f_1$ and $f_2$ within a certain range, this tachyonic mode can be lifted above the BF bound \cite{Bergman:2010gm}.

For this spatially homogeneous ansatz, the D7-brane action has a DBI term given by
\bea
\label{DBI_action}
 S_{DBI} & = & -T_7 \int d^8x\, e^{-\Phi} \sqrt{-\mbox{det}(g_{\mu\nu}+ 2\pi\alpha' F_{\mu\nu})} \nonumber \\
 &=& - {\cal N} \int dr\, r^2\sqrt{\left(4\cos^4\psi + f_1^2 \right)
 \left(4\sin^4\psi + f_2^2 \right)}\times \nonumber \\ 
 & & \qquad\qquad \times \sqrt{\left(1+ r^4 h z'^2+ r^2 h \psi'^2 - {a_t'}^2\right)\left(1+\frac{b^2}{r^4}\right)} \,,
\eea
and a CS term given by
\bea
\label{CS_action}
S_{CS} &=& -\frac{(2\pi\alpha')^2T_7}{2} \int P[C_4]\wedge F \wedge F \nonumber\\
&=&  -{\cal N}f_1 f_2 \int dr\, r^4 z'(r)
+ 2{\cal N} \int dr\, c(r) b a_t'(r) \ ,
\eea
where ${\cal N} \equiv 4\pi^2 L^5 T_7 V_{3}$, the volume of the $(2+1)$-dimensional defect is $V_3$, and 
\be
c(r) = \psi(r) - \frac{1}{4}\sin\left(4\psi(r)\right) - \psi_\infty + \frac{1}{4}\sin(4\psi_\infty) \ .
\label{cr}
\ee
Note that $c(r)$, and therefore $\psi(r)$, plays the role of an axion. The equations of motion derived from this action, first derived in \cite{Bergman:2010gm}, are given in Appendix \ref{app:backgroundeom}. 

The asymptotic behavior of the fields is given by
\bea
\psi(r) & \sim & \psi_{\infty}+mr^{\Delta_{+}}-c_{\psi}r^{\Delta_{-}} \\ 
z(r) &\sim& z_{0} +\frac{f_1 f_2}{r} \frac{1}{ \sqrt{ (f_1^2+4\cos^4\psi_\infty)(f_2^2+4\sin^4\psi_\infty)-f_1^2 f_2^2 }} \\ 
a_t(r) &\sim&  \mu - \frac{d}{r} \frac{1}{ \sqrt{ (f_1^2+4\cos^4\psi_\infty)(f_2^2+4\sin^4\psi_\infty)-f_1^2 f_2^2 }}\, \label{ddefintion}, 
\eea
where the boundary value $\psi_\infty$ and the exponents $\Delta_\pm$ are fixed by the 
fluxes $f_1$ and $f_2$:
\bea
(f_1^2 + 4\cos^4\psi_\infty)\sin^2\psi_\infty =  (f_2^2 + 4\sin^4\psi_\infty)\cos^2\psi_\infty \label{const_solution} \\[5pt]
\Delta_\pm  =  -\frac{3}{2}\pm \frac{1}{2}\sqrt{9+16\frac{f_1^2
+16\cos^6\psi_\infty-12\cos^4\psi_\infty}{f_1^2+4\cos^6\psi_\infty}} \,. \label{deltapm}
\eea
The parameters $m$ and $c_\psi$ correspond to the ``mass" and ``condensate" of the fundamental fermions, respectively, and $\mu$ and $d$ to the chemical potential and charge density, respectively. The physical charge density, defined by the variation of the on-shell action with respect to the boundary value of $A_t$, is given by $D=\frac{2\pi\alpha'\mathcal N}{LV_3} d$.

In this paper, we are concerned exclusively with black hole embeddings, for which the D7-brane probe reaches the horizon at $r=r_T$ and which correspond to a gapless state.  And, for simplicity, we focus only on massless $m=0$ embeddings.  These embeddings are trivial,  in the sense that $\psi$ is constant, if either $d$ or $b$ are zero.

 The properties of this phase, including the anomalous Hall conductivity and spontaneous magnetization, have been extensively studied \cite{Bergman:2010gm, Bergman:2011rf}.  In particular, at high density and low temperature and magnetic field, the homogeneous black hole embedding suffers from an instability at nonzero momentum \cite{Bergman:2011rf, Jokela:2012vn}, which results in a spatially modulated ground state \cite{Jokela:2014dba}.

In the following sections, when we present specific numerical results in figures, we choose the specific fluxes $f_1 =  f_2=\frac{1}{\sqrt 2}$, such that $\Delta_+=-1$ and $\Delta_-=-2$ and thus $\psi_\infty=\frac{\pi}{4}$. 
We will keep the formulas in their general form, however.

\subsection{Anyonization}
\label{sec:anyonization}

Given any $(2+1)$-dimensional CFT with a conserved U(1) current $J_\mu$, there are two natural transformations mapping it into another CFT: adding a Chern-Simons term for an external vector $\mathcal B_\mu$ coupled to $J_\mu$ and turning that external vector into a dynamical field \cite{Witten:2003ya, BurgessandDolan}.  Combining these operations generates an $SL(2,{\mathbb{R}})$ which acts via
\bea
{J}^{*}_{\mu}&=&a_sJ_{\mu}+b_s {\cal B}_{\mu} \nonumber \\
{\cal B}^{*}_{\mu}&=&c_{s} J_{\mu}+d_{s}  {\cal B}_{\mu} \ .
\label{SL2Z}
\eea
Charge quantization restricts the transformations to $SL(2,{\mathbb{Z}})$.

The action of this $SL(2,{\mathbb{Z}})$ transformation can be understood in two ways.  
One interpretation is that the mapping yields a new theory with current ${J}^{*}_{\mu}$ and magnetic field ${\cal B}^{*}_{\mu}$.  The particles carrying this current have anyon statistics, with a statistical angle\footnote{It has been argued that the statistical angle $\theta$ should be quantized \cite{Wen}.  However, \cite{Fitzpatrick:2012ww} has also suggested this argument may not hold for gapless anyons.} 
\be
\label{statistical_angle}
\theta = \pi \left(1- \frac{c_s}{d_s} \right) \ .
\ee

An alternative view of the new theory is as a description of the original particles and fields but with different dynamics.  The original theory describes the behavior of the current $J_\mu$ in presence of a non-dynamical background field $\mathcal B_\mu$.  In the transformed theory, the dynamical variable is the combination $a_s J_\mu + b_s \mathcal B_\mu$, with $c_s J_\mu + d_s \mathcal B_\mu$ held fixed.  

In general, holding fixed a linear combination of the current and the field might seem somewhat strange.  However, for the S-transformed  theory, where $a_s = d_s = 0$, $c_s = - b_s = 1$,
\bea
{J}^{*}_{\mu}&=&- {\cal B}_{\mu} \nonumber \\
{\cal B}^{*}_{\mu}&=&J_{\mu} \ ;
\eea
the roles of the dynamical current and background field are exchanged.\footnote{Note that for the S-transformed theory, the flux attachment picture of an anyon current breaks down; $d_s = 0$ implies an infinite amount of attached flux and, via equation \eqref{statistical_angle}, an infinite statistical angle.} The new theory describes a dynamical $U(1)$ gauge field  coupled to a fixed charged current.  Although this gauge field is dynamical,  i.e., integrated in the path integral, it does not have a Maxwell term in the action.  As we will explain in Sec.~\ref{sec:2+1QED}, this S-transformed theory is then a good approximation of $(2+1)$-dimensional QED, either at very large coupling or with a very large number of fermionic flavors.

Holographically, the background field $\mathcal B$ is dual to the boundary value of the bulk gauge field $A_\mu$
\be
{\cal B}_{\mu} = \frac{1}{2\pi}\epsilon_{\mu \rho \nu}\partial^{\rho}A^{\nu}\Big|_\partial  \ ,
\ee
while $J_\mu$ is given by the variation of the on-shell bulk action with respect to the boundary value of $A_\mu$
\be
J_\mu = \frac{\delta S}{\delta A^\mu}\Bigg|_\partial,
\ee
which, by the equations of motion, is proportional to the radial derivative $\partial_r A_\mu$. 

In the bulk, the $SL(2,{\mathbb{Z}})$ transformation acts on the asymptotic boundary conditions for $A_\mu$.  In the original CFT, the external field $\mathcal B_\mu$, and therefore the boundary value of $A_\mu$, is held fixed.  After the transformation \eqref{SL2Z}, fixing the new external field $\mathcal B^*_\mu$ translates in the bulk to fixing a linear combination of $A_\mu$ and $\partial_r A_\mu$ at the boundary:
\be
\label{boundary_condition_c_over_d}
0 = \frac{\delta B*}{2\pi} = - c_s \delta D + d_s \frac{\delta B}{2\pi} \ .
\ee
This mixed boundary condition on $A_\mu$ corresponds to alternative quantization of the bulk field. 

The bulk equations of motion are unaffected by the transformation \eqref{SL2Z}, and bulk solutions therefore remain solutions under $SL(2,{\mathbb{Z}})$, even though their holographic interpretation is altered. However, as explained in the following subsection, changing the boundary conditions does significantly affect the spectrum of fluctuations around these bulk solutions.  For example, the Minkowski embedding holographically dual to a gapped quantum Hall state in the standard quantization can be transformed by an appropriate $SL(2,{\mathbb{Z}})$ mapping into a gapless superfluid \cite{Jokela:2013hta}.

\subsection{Fluctuations}
\label{sec:fluctuations}

In this section, we set up the fluctuation analysis of the homogeneous anyon fluid described by the BH embedding of the D3-D7' system.  We pay particular attention to how the anyon statistics change the boundary conditions for bulk excitations and thereby impact the stability of the homogeneous fluid.

We have the freedom to scale out one parameter of the model, and for this we choose the temperature. We introduce a new compact radial variable
\be
 u = \frac{r_T}{r} \ .
 \ee
The boundary is located at $u=0$, and the horizon is at $u=1$. To scale out the dependence on $r_T$, we introduce notation with hats as follows:
\be
 \hat x^\mu = r_T x^\mu  \quad ,\quad \hat z= r_T z  \quad , \quad \hat a_\mu = \frac{a_\mu}{r_T} \ .
\ee
The rescaled density and magnetic field are then
\be
\hat d = \frac{d}{r_T^2} \quad , \quad
\hat b = \frac{b}{r_T^2} \ .
\ee

We consider excitations of the D7-brane gauge field $\hat a_\mu$ and embedding scalars $\psi$ and $\hat z$ that are constant on the internal $S^5$. Rotational invariance of the homogeneous state allows us to align the direction of the fluctuations with the $x$-axis.  Working in radial gauge $\hat a_u = 0$, the fluctuations take the form:
\bea
\delta \psi & = & \delta \tilde \psi(u) e^{-i\omega t + ikx} \\
\delta \hat z_\mu & = & \delta \tilde z_\mu(u) e^{-i\omega t + ikx} \\
\delta \hat a_\mu & = & \delta \tilde a_\mu(u) e^{-i\omega t + ikx} \ .
\eea
The frequency and momentum can also be rescaled by the horizon radius:
\be
 \hat \omega  = \frac{\omega}{r_T} \quad ,\quad \hat k_\mu= \frac{k_\mu}{r_T} \ .
\ee
It is useful to work in terms of the gauge-invariant combination:
\be
\delta\tilde e_x = \hat\omega \delta \tilde a_x + \hat k \delta \tilde a_t \ .
\ee

We now expand the D7-brane action \eqref{DBI_action} and \eqref{CS_action}  to second order in the fluctuations and derive the linearized equations of motion for $\delta\tilde \psi$, $\delta\tilde z$, $\delta\tilde e_x$, and $\delta\tilde a_y$.  In addition, the equation of motion coming from the variation of $\delta \hat a_u$ provides a constraint enforcing the gauge condition $\hat a_u = 0$. This rather lengthy coupled system of equations can be found in Appendix \ref{app:fluctuationeom}.   All the fluctuations $\delta\tilde \psi$, $\delta\tilde z$, $\delta\tilde e_x$, and $\delta\tilde a_y$ are in general coupled.

As explained above, the $SL(2,{\mathbb{Z}})$ operation which transforms the statistics also mixes the background field and conserved current.  This corresponds to changing the boundary conditions obeyed by the holographic dual fields.  Initially, the background field ${\cal B}$ is fixed, which translates to Dirichlet boundary conditions for the bulk gauge field $\delta a_\mu\big|_\partial = 0$. After applying an $SL(2,{\mathbb{Z}})$ transformation, $c_s \delta J_\mu + d_s \delta {\cal B}_\mu = 0$ instead.  From the variation of the action \eqref{DBI_action}, the conserved current is given by
\be
J_\mu  = \frac{2\pi\alpha'\mathcal N}{LV_3}  \sqrt{\left( f_1^2 + 4\cos^4\psi_\infty \right)\left(f_2^2 + 4\sin^4\psi_\infty \right) - f_1^2f_2^2} \ \partial_u \hat a_\mu \ .
\ee

Instead of using the $SL(2,{\mathbb{Z}})$ parameters $c_s$ and $d_s$, as in equation \eqref{boundary_condition_c_over_d}, it is convenient to parameterize the general alternative boundary condition in the following way
\be
\label{alternativeboundarycondition}
-n \delta F_{\mu u} + \frac{1}{2} \epsilon_{\mu\nu\rho} \delta F^{\nu\rho} = 0 \ ,
\ee
where $n$ is given by
\be
\label{n_def}
n = \frac{c_s}{d_s} \frac{N_c}{\pi} \sqrt{\left( f_1^2 + 4\cos^4\psi_\infty \right)\left(f_2^2 + 4\sin^4\psi_\infty \right) - f_1^2f_2^2} \ . 
\ee  
The parameter $n$ controls the degree to which the anyons differ from fermions; $n=0$ gives the original Dirichlet boundary condition, yielding a fluid of fermions.  Note that for order one $c_s/d_s$, the parameter $n$ is order $N_c$, and the statistical angle \eqref{statistical_angle} is then $\theta = \pi - \mathcal O(1/N_c)$.
Also, note that the parameter $n$ does not completely specify the $SL(2,{\mathbb{Z}})$ transformation.  In general, an $SL(2,{\mathbb{Z}})$ transformation which changes $n$ also affects the charge and magnetic field.  However, the anyonic charge density $\hat d^*$ and magnetic field $\hat b^*$ after the transformation will depend on the values of $a_s$, $b_s$, $c_s$ and $d_s$ in equation \eqref{SL2Z}.

Using the boundary limit $u \to 0$ of the gauge constraint coming from the $\delta \hat a_u$ equation of motion \eqref{deltaauEOM},
\be
\hat k\delta \partial_u \tilde a_x + \hat\omega \delta\partial_u \tilde a_t = 0 \ ,
\ee
the boundary condition \eqref{alternativeboundarycondition} can be written as
\bea
\label{alternativeboundarycondition1}
\frac{i n}{\hat\omega^2 - \hat k^2} \delta \partial_u \tilde e_x + \delta \tilde a_y=0 \\
\label{alternativeboundarycondition2}
 -in \delta \partial_u \tilde a_y + \delta \tilde e_x=0 \ .
\eea

Our goal is to find the spectrum of quasinormal modes (QNM) as a function of $n$.  These correspond to values of $(\hat\omega, \hat k)$ for which there are normalizable solutions of the coupled fluctuation equations of motion (\ref{deltapsiEOM}), (\ref{deltazEOM}), (\ref{deltaatEOM}), (\ref{deltaaxEOM}), (\ref{deltaayEOM}) with the boundary conditions \eqref{alternativeboundarycondition1} and \eqref{alternativeboundarycondition2}.  We search for such solutions numerically, using the techniques described in \cite{Bergman:2011rf, Jokela:2012vn} and pioneered in \cite{Kaminski:2009dh, Amado:2009ts}.


\section{Anyonic Stripes}
\label{sec:anyonicstripes}

\begin{figure}[ht]
\centering
\includegraphics[width=0.7\textwidth]{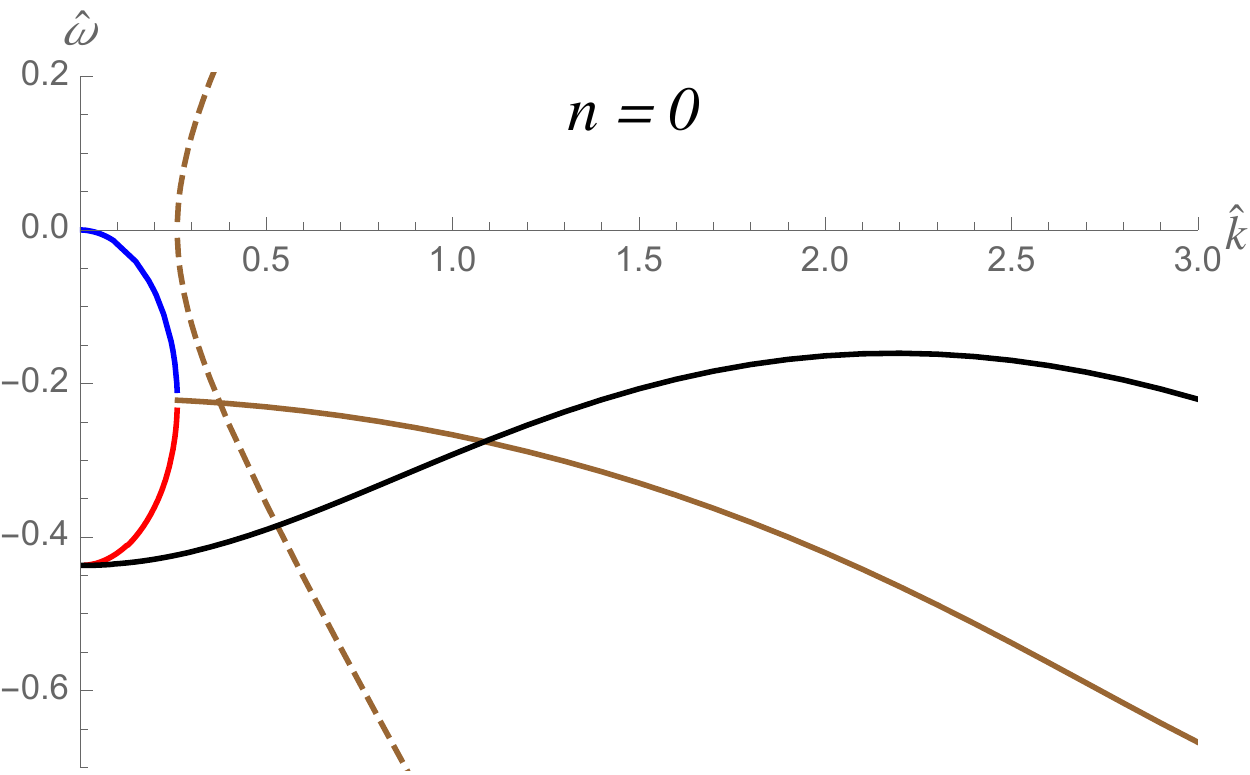} 
\caption{The quasinormal spectrum for fixed $\hat d=4$, $\hat b = 0$, and $n =0$.  Solid curves show ${\rm Im}\ \hat\omega$, and dotted curves give ${\rm Re}\ \hat\omega$.  The hydrodynamical mode is shown in blue, the next-lowest imaginary longitudinal mode is red, the lowest imaginary transverse mode is black, and the complex holographic zero sound mode is brown.}
\label{fig:n=0} 
\end{figure}

Let us first review the QNM spectrum for the fermionic system, with the original boundary conditions $n=0$ and zero magnetic field $\hat b=0$, studied in \cite{Bergman:2011rf}.  A representative example is shown in Fig.~\ref{fig:n=0}. At long wavelength $\hat k \to 0$, the lowest QNM is a longitudinal hydrodynamical mode associated with charge conservation:
\be
\hat\omega = -i \hat D \hat k^2 + \dots
\ee
where $\hat D$ is the rescaled diffusion constant.  The next-longest-lived QNMs are a pair of purely imaginary modes, one longitudinal and one transverse, which have the same frequency at $\hat k=0$ due to the restoration of rotational invariance there. 

As $\hat k$ is increased the hydrodynamical mode eventually merges with this purely imaginary longitudinal mode. These combine into two complex modes which are identified as the holographic zero sound modes, propagating in opposite directions.  This transition represents the crossover from the hydrodynamic to the collisionless regime. 

The longest lived transverse mode, the only purely imaginary mode in the collisionless regime, is responsible for the spatially modulated instability.  As $\hat k$ is increased, ${\rm Im} \ \hat\omega$ initially increases also, reaching a local maximum at some nonzero momentum.  At a critical density, $\hat d_{cr} \approx 5.5$, this maximum enters the upper half of the complex $\omega$-plane, signaling that fluctuations with $\hat k_{cr} \approx 2.8$ have become unstable.  At this point, there is a second-order phase transition to a striped phase with spatially modulated charge and spin densities as well as a modulated transverse current\cite{Jokela:2014dba}.  Just above the critical point, the spatial frequency of these stripes $\hat k_0 = \hat k_{cr}$ and then gradually increases with increasing $\hat d$.

Now we investigate how changing the statistics impacts the QNM spectrum.  As we do this, the anyon charge density $\hat d^*$ and background magnetic field $\hat b^*$ are given by equation \eqref{SL2Z} in terms of the original charge density $\hat d$, magnetic field $\hat b$, and the parameters $a_s$, $b_s$, $c_s$, and $d_s$ of the $SL(2,{\mathbb{Z}})$ transformation.  However, the boundary conditions of the bulk gauge field, and therefore the QNM spectrum, depend only on $n$, which is proportional to $c_s/d_s$.   
For this reason, it is much more convenient to label the solutions and the QNM spectra by $n$, $\hat d$, and $\hat b$. 
Note that varying $n$ for fixed $\hat d$ and $\hat b$ changes not only the anyon statistics but also, in general, $\hat d^*$ and $\hat b^*$.\footnote{For a given $n$, one could attempt to choose the $SL(2,{\mathbb{Z}})$ parameters such that $\hat d^* = \hat d$ and $\hat b^* = \hat b$, but because $a_s$, $b_s$, $c_s$, and $d_s \in \mathbb{Z}$ this is, in general, not possible.}

\begin{figure}[ht]
\centering
\includegraphics[width=0.42\textwidth]{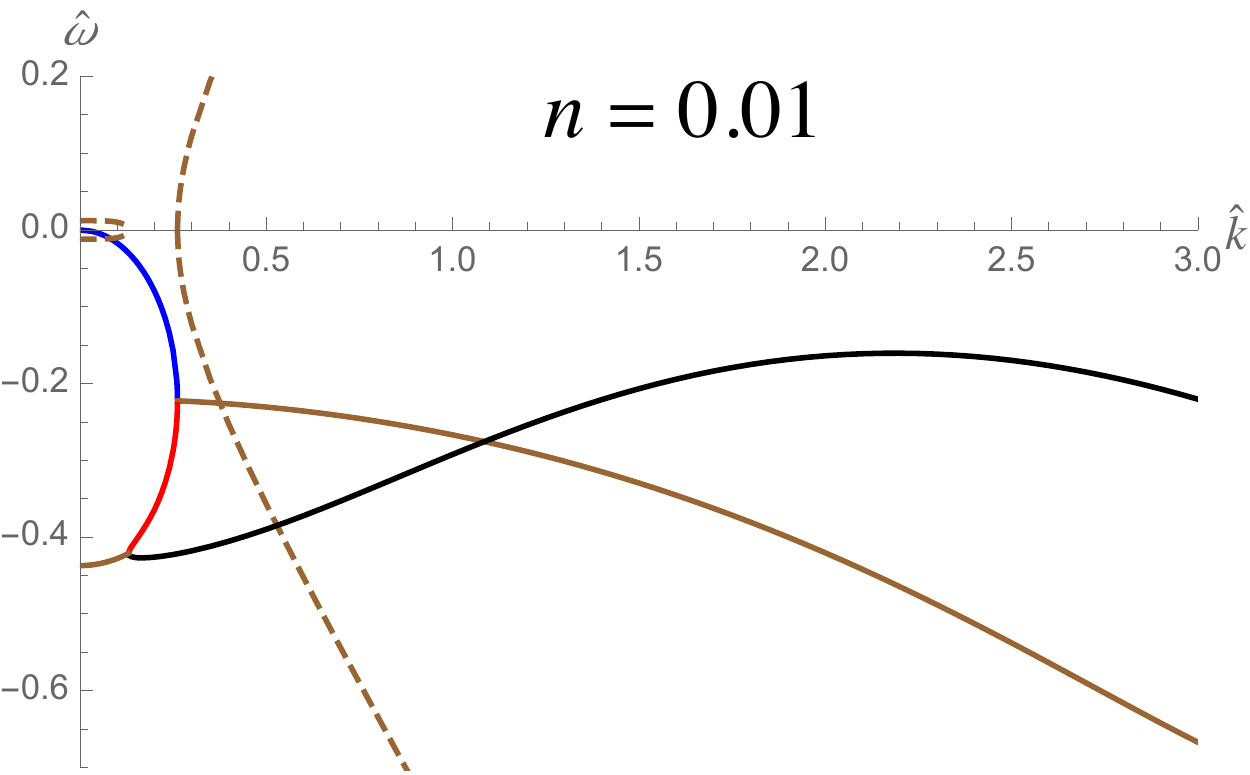}  
\includegraphics[width=0.42\textwidth]{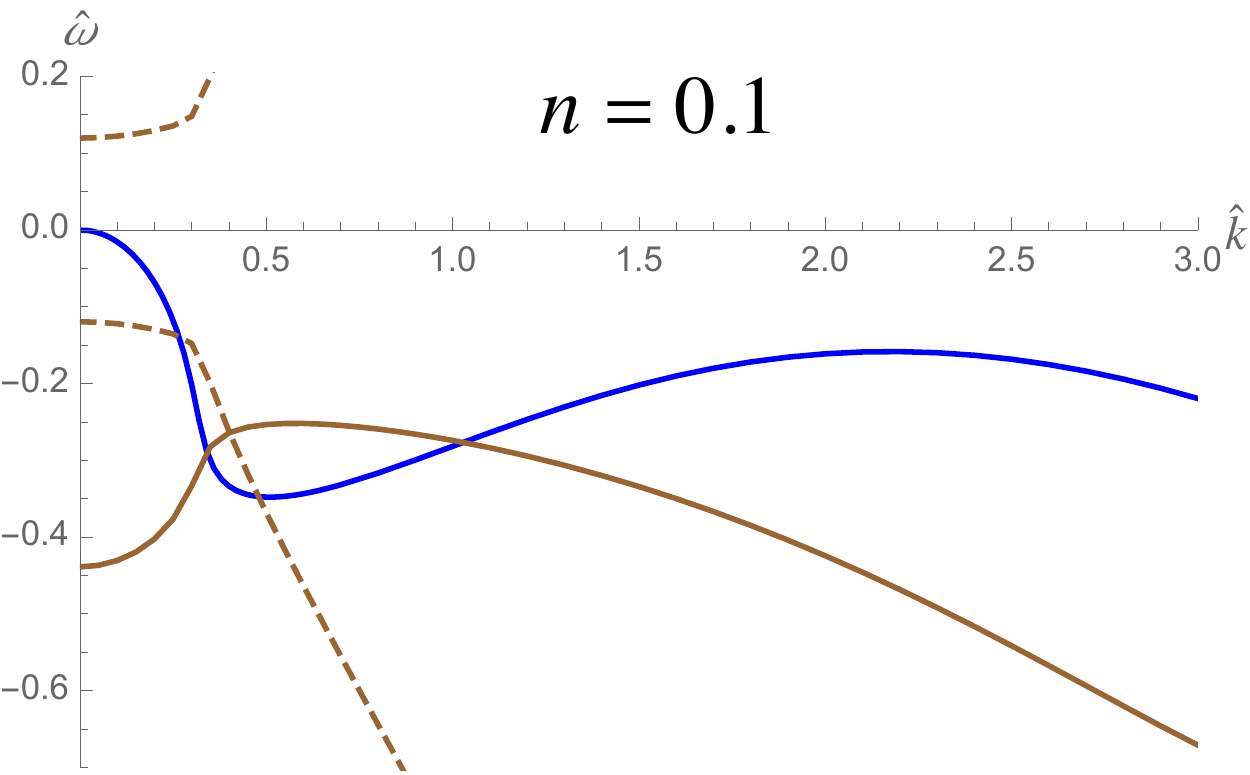} 
\includegraphics[width=0.42\textwidth]{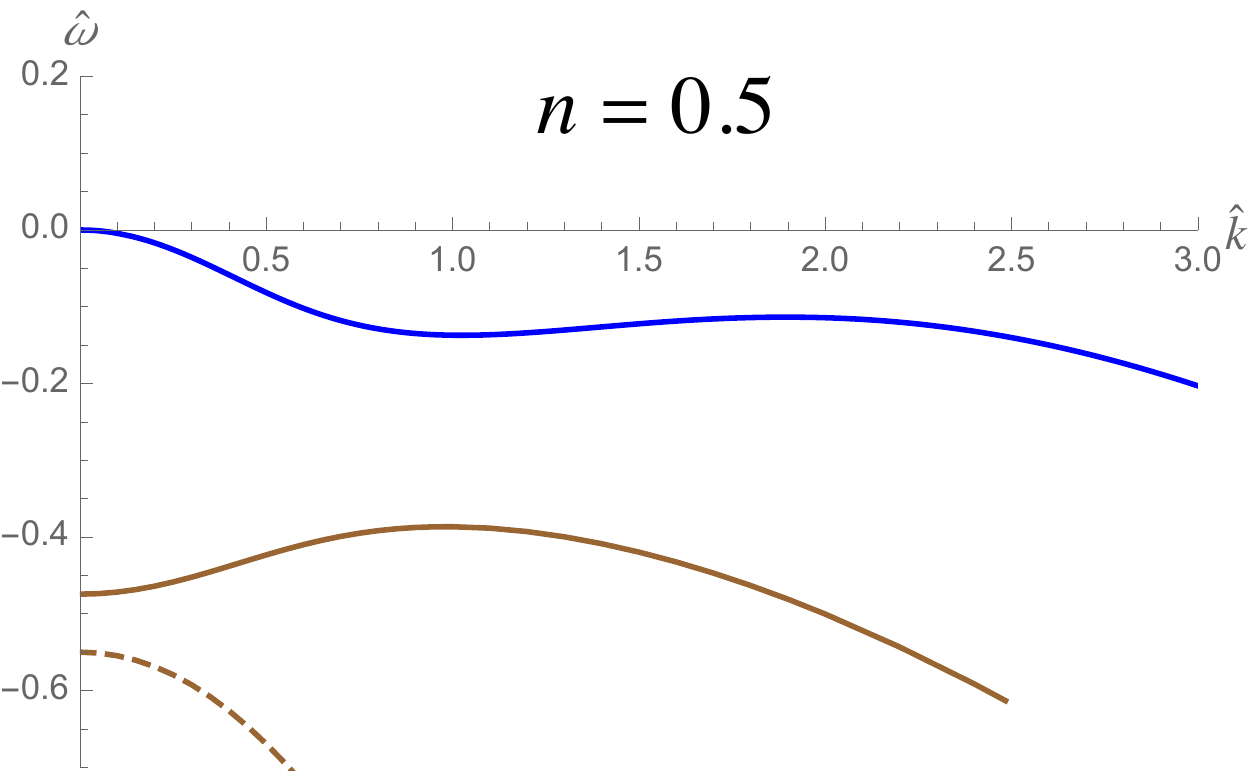}
\includegraphics[width=0.42\textwidth]{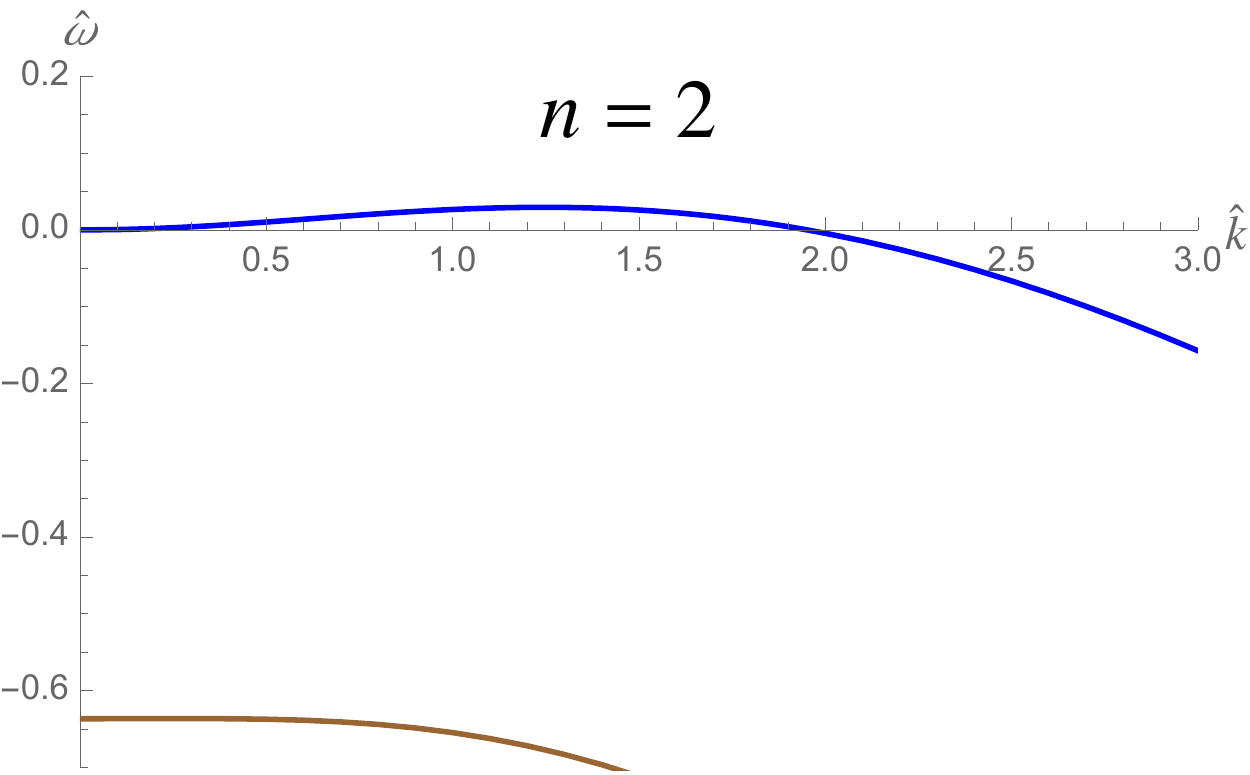}
\caption{The quasinormal spectrum for fixed $\hat d=4$ and $\hat b = 0$ and varying $n$: (upper left) $n=0.01$, (upper right) $n=0.1$,  (lower left) $n=0.5$, and (lower right) $n=2$.  Solid curves show ${\rm Im}\ \hat \omega$, and dotted curves give ${\rm Re}\ \hat \omega$.  The hydrodynamical mode is shown in blue, the next lowest imaginary longitudinal mode is red, the lowest imaginary transverse mode is black, and the complex holographic zero sound mode is brown.}
\label{fig:change_n} 
\end{figure}

We begin by analyzing the effect of increasing $n$, keeping $\hat b = 0$; see Fig.~\ref{fig:change_n}.  For fixed $\hat d$, as we increase $n$ from zero, longitudinal and transverse modes mix; at very small $\hat k$, instead of two imaginary modes, the next-longest-lived QNMs are combined into two complex modes.  For sufficiently small $n$, at some nonzero $\hat k$, they split into two purely imaginary modes.  Then, as in the $n=0$ case, the longitudinal mode then combines with the hydrodynamical mode at larger $\hat k$ to form the holographic zero sound, and the transverse mode is the lowest purely imaginary mode at large $\hat k$.

However, as $n$ is increased further, the mixing is stronger, and the longitudinal and transverse modes never split into purely imaginary modes again.  The holographic zero sound then extends to $\hat k = 0$, but with a mass gap. In this case, the hydrodynamical mode, without another imaginary mode with which to merge, continues to large $\hat k$, and there is no crossover to a collisionless regime. 

The choice of statistics also impacts the nature of the modulated instability, both the critical $\hat d$ at which it occurs as well as the mode responsible.  Fig.~\ref{fig:change_n} shows that, for $\hat d = 4$, the imaginary frequency of the hydrodynamical mode increases with $n$.  At $n \approx 2$, the mode crosses the ${\rm Im} \  \hat \omega$ axis and becomes unstable.

\begin{figure}[!ht]
\centering
\includegraphics[width=0.42\textwidth]{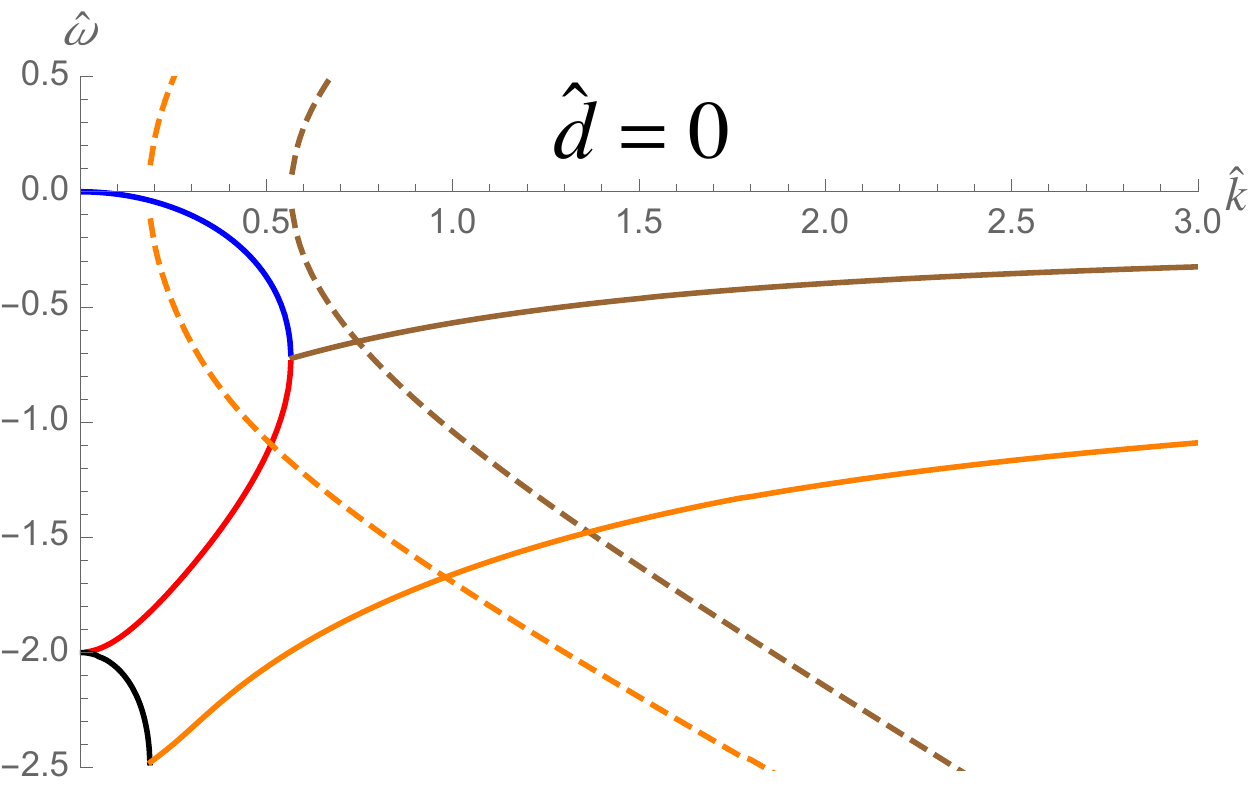} 
\includegraphics[width=0.42\textwidth]{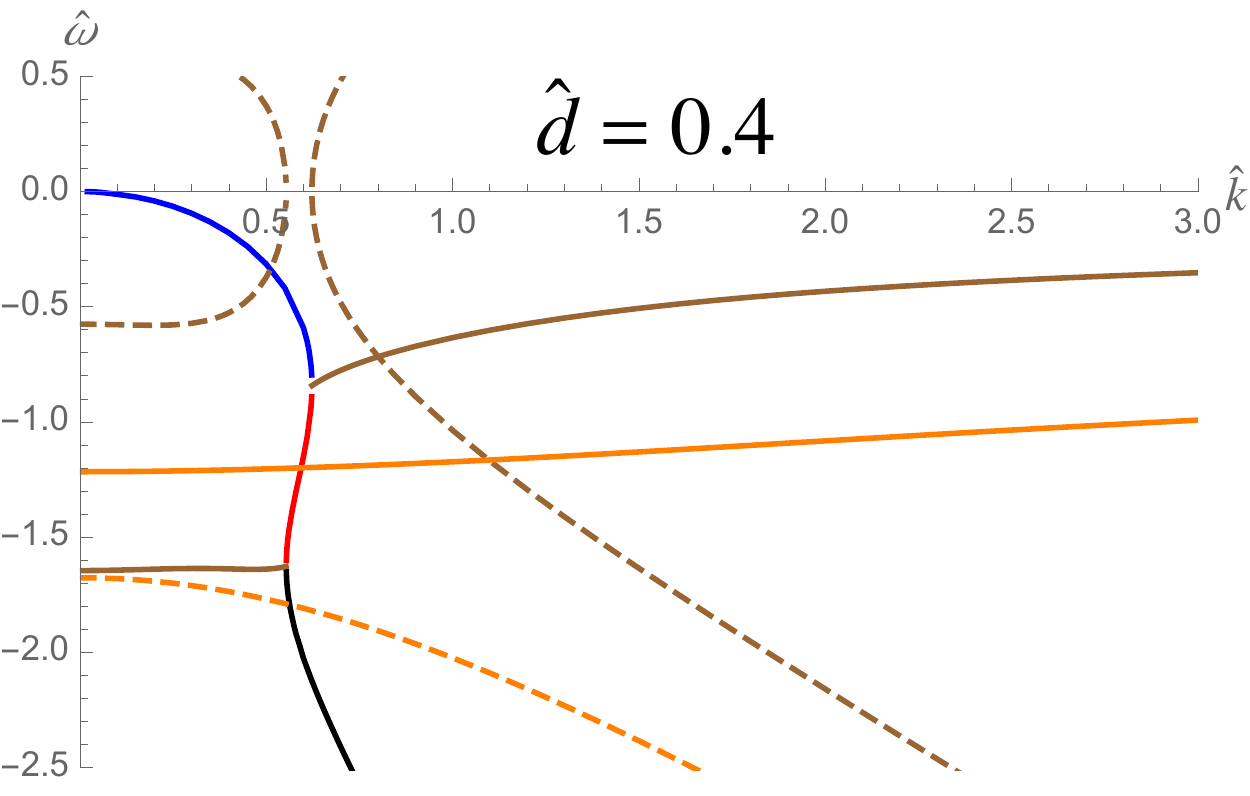}
\includegraphics[width=0.42\textwidth]{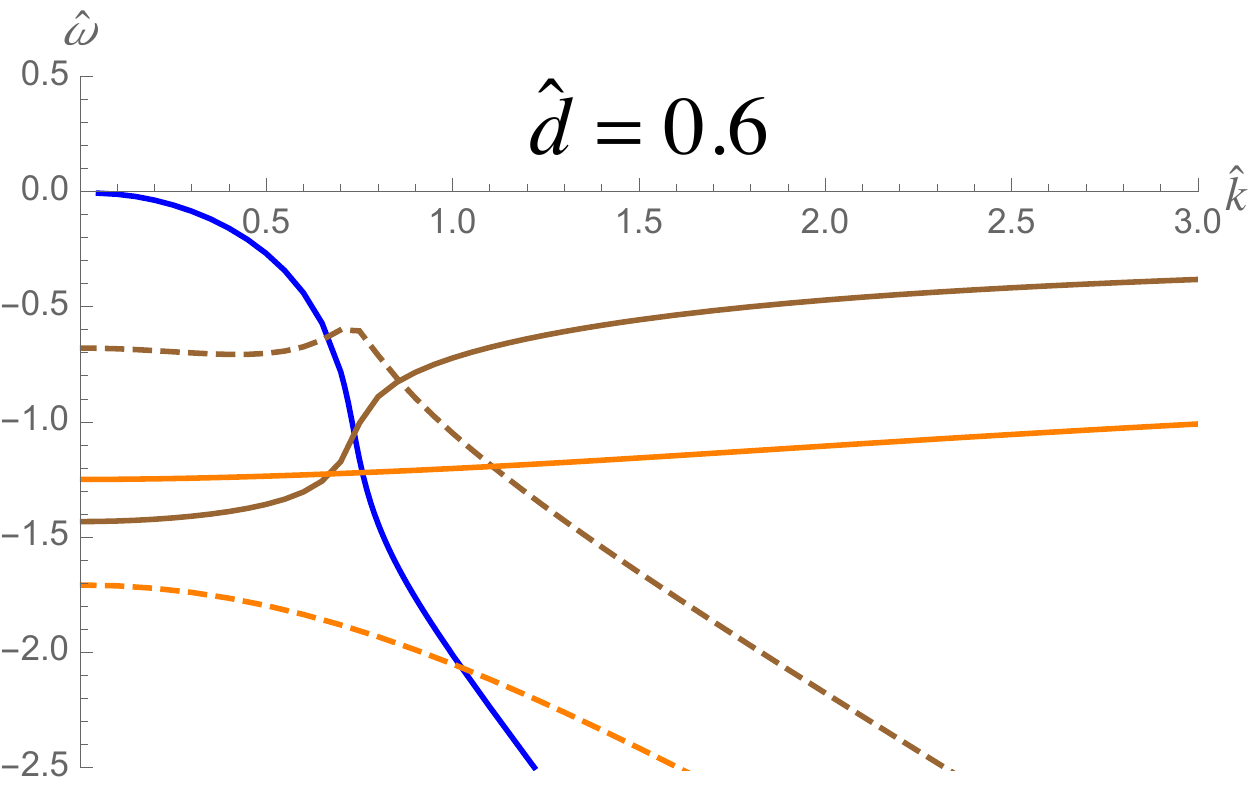}
\includegraphics[width=0.42\textwidth]{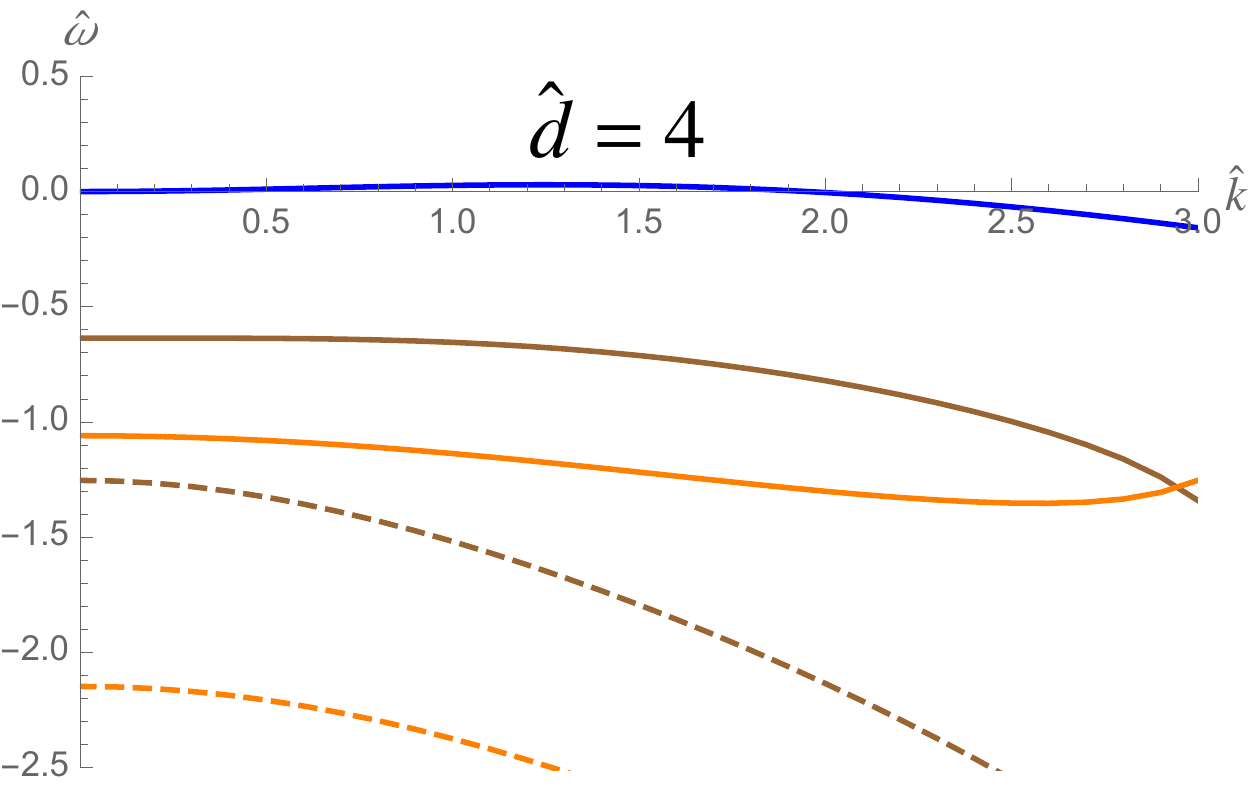}
\caption{The quasinormal mode spectrum for $n=2$ and $\hat b = 0$ and varying $\hat d$: (upper left) $\hat  d=0$ (upper right) $\hat d = 0.4$, (lower left) $\hat d = 0.6$, and (lower right) $\hat d = 4$.   As in Fig.~\ref{fig:change_n} above, solid curves show ${\rm Im}\ \hat \omega$, and dotted curves give ${\rm Re}\ \hat\omega$.  The hydrodynamical mode is again shown in blue, the next-lowest imaginary longitudinal mode is red, the lowest imaginary transverse mode is black, and the complex holographic zero sound mode is brown.  Another complex mode is shown in orange.}
\label{fig:change_d} 
\end{figure}

This process can also be illustrated by varying $\hat d$ for a fixed choice of $n$. 
As we increase $\hat d$, the spectrum evolves in a similar way as above; see Fig.~\ref{fig:change_d}.  For $\hat d=0$ and $n > 0$, the spectrum is qualitatively similar to that of $n=0$ and $\hat d > 0$.  The hydrodynamical mode and the next-lowest, purely imaginary longitudinal mode merge at some nonzero $\hat k$ to form a complex holographic zero sound.  As $\hat d$ is increased, the mixing at small $\hat k$ between the non-hydrodynamic longitudinal mode and the lowest transverse modes gets stronger, resulting in a massive holographic zero sound.

At this point, the hydrodynamic mode then no longer merges with another mode and persists to large $\hat k$, with an imaginary frequency which increases with $\hat d$. Eventually, at a critical  $\hat d_{cr}$, ${\rm Im} \ \hat \omega$ becomes positive for modes with momentum $\hat k_{cr}$, and the homogeneous phase becomes unstable.

For $n \gsim 0.1$, the unstable mode is continuously connected to the hydrodynamic charge diffusion mode.  The instability is then an example of uphill diffusion, where the charges spontaneously separate into positive and negative charge regions. 

\begin{figure}[!ht]
\centering
\includegraphics[width=0.45\textwidth]{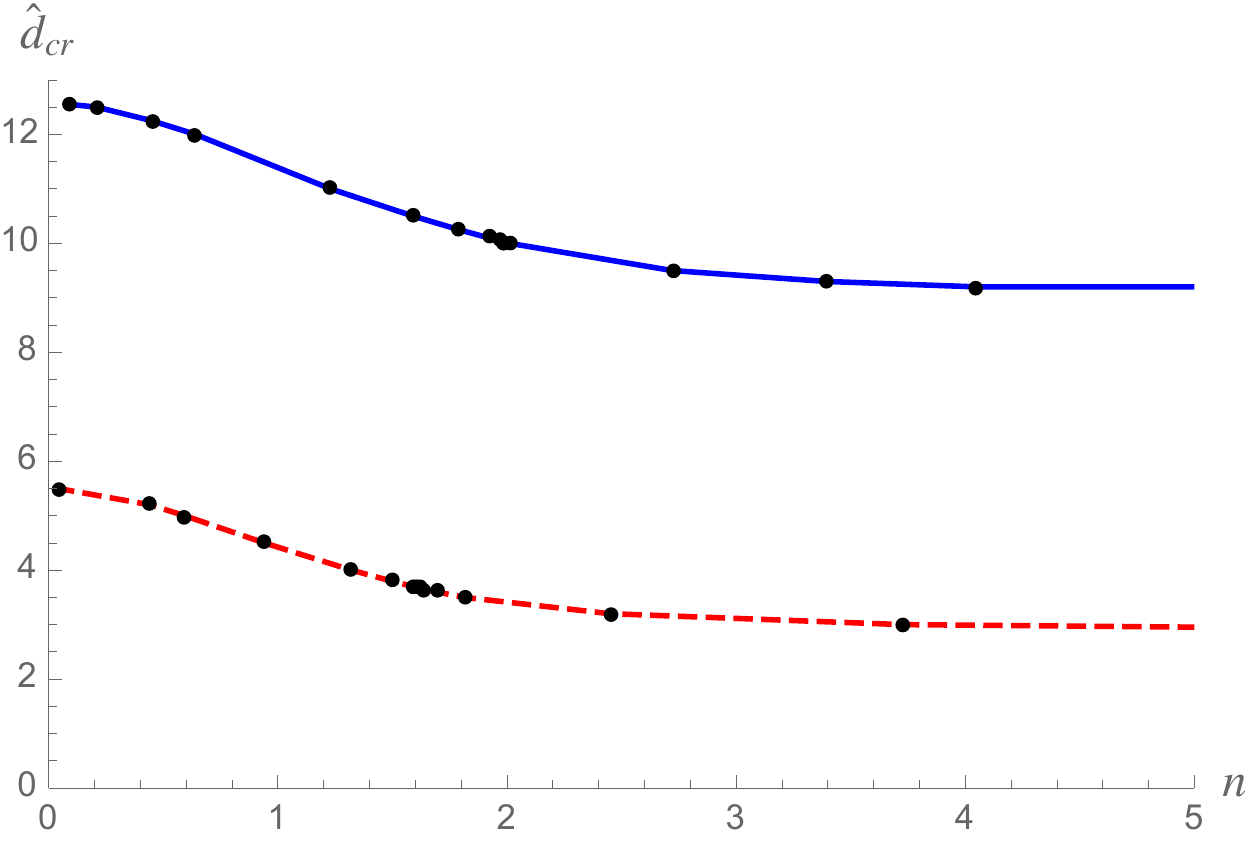} 
\includegraphics[width=0.45\textwidth]{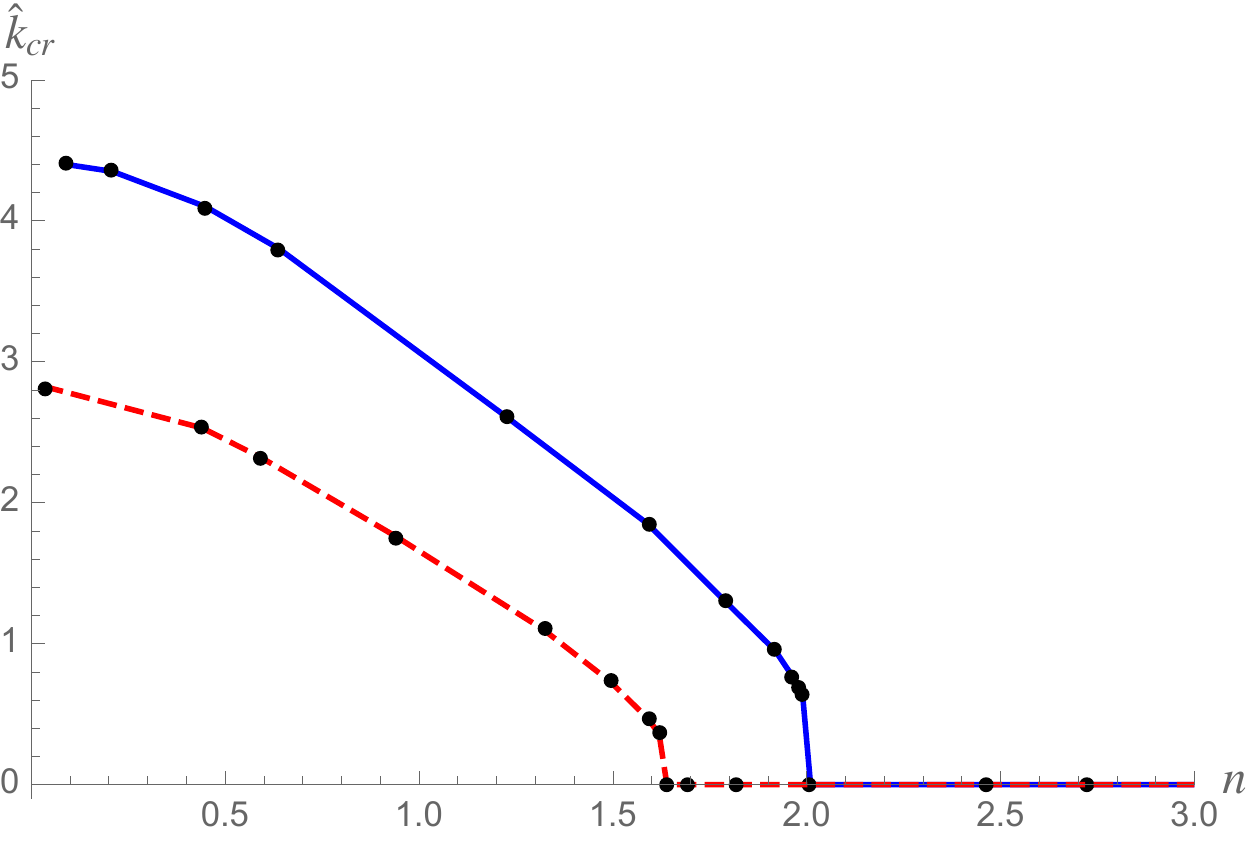}
\caption{(Left) The critical  $\hat d_{cr}$ as a function of $n$, and (right) the critical momentum $\hat k_{cr}$ vs $n$.  For both figures, the lower, dashed red curve shows $\hat b=0$ while the upper, solid blue curve is $\hat b=1$.  The curves are fit to the numerically computed black data points.}
\label{fig:dkcr} 
\end{figure}

Computing the onset of the instability as a function of $n$, we see that both $\hat d_{cr}$ and $\hat k_{cr}$ depend on the statistics; both decrease with increasing $n$. This behavior is illustrated in Fig.~\ref{fig:dkcr}. The decrease in $\hat d_{cr}$ implies that increasing $n$ enhances the instability.

Of particular note is that $\hat k_{cr}$ is zero above some finite value of $n$. What is the physical meaning of $\hat k_{cr}$ going to zero?  Let us prepare the system with small enough  $\hat d$ such that we are in the homogeneous phase.  If $\hat d$ is then increased to $\hat d_{cr}$, a phase transition to a modulated phase is expected with stripes of width $\sim \frac{2 \pi}{\hat k_{cr}}$. As $\hat k_{cr} \rightarrow 0$, the stripes grow in size, implying that, for large enough $n$, the modulated phase is just a separation of space into two halves.

However, to properly determine the nature of the modulated ground state, we need to go beyond the fluctuation analysis and solve for the nonlinear inhomogeneous solution. This has to be done for general $n$ in the same way the striped ground state was computed previously for $n=0$ in \cite{Jokela:2014dba}.  The presence of an instability at nonzero $\hat k$ implies a phase transition to a spatially modulated phase, and, for $n=0$, the expected striped ground state was indeed realized.  To compute the dynamically preferred wavelength of the stripes, the free energy must be minimized over different values of $\hat k$.  However, computing the free energy for general $n$ is quite subtle.  Implementing alternative boundary conditions on the bulk gauge field induces several terms into the action which need to be properly taken into account when computing the free energy \cite{Ihl:2016sop}.

Without currently having access to the ground state at generic $n$, we do not know for certain whether the instability with $\hat k_{cr} =  0$ really leads to a phase separation or to a spatial modulation with finite wavelength. Features of the unstable mode imply similar features of the ground state to which the instability leads, but the implication is by no means guaranteed.  

One exception is $n=\infty$, for which we can readily compute the free energy and explicitly solve for the modulated ground state, which we will discuss further in Sec.~\ref{sec:2+1QED} below.
As we will show, the preferred ground state for $\hat{d} > \hat{d}_{cr}$ is actually spatially modulated with finite wavelength.  For $n = \infty$, we find that the phase transition is continuous, as was found for $n=0$ in \cite{Jokela:2014dba}.  In both cases, as the transition is approached from the striped phase, $\hat{d} -  \hat{d}_{cr} \to 0^+$, the amplitude of the modulation also goes to zero.  For $n=\infty$, the frequency of the spatial modulation also continuously goes to zero at the critical point, while for $n=0$, it stays nonzero.  The behavior of $\hat k_{cr}$ in Fig.~\ref{fig:dkcr} strongly suggests a transition between these two types of critical behavior around $n \approx 1.6$.

We have so far kept $\hat b = 0$; now, we turn it back on. The QNM spectrum with the standard quantization $n=0$ in the presence of a magnetic field was thoroughly studied in \cite{Jokela:2012vn}, and two notable effects were found: A nonzero $\hat b$ mixes the modes and, at large enough $\hat b$, leads to a massive holographic zero sound.  This behavior is very similar to the mixing caused by changing the statistics $n$, seen in Fig.~\ref{fig:change_n}.  However, unlike the anyon statistics,  $\hat b$ suppresses the modulated instability; the critical $\hat d_{cr}$ increases with $\hat b$.  In addition, it was shown in \cite{Jokela:2014dba} that at nonzero $\hat b$, the phase transition to the modulated phase becomes first order. 

Investigating the impact of increasing $\hat b$ with $n \not= 0$, we find that anyon statistics do not notably alter these effects.
 $\hat b$ still enhances mode mixing and mitigates the modulated instability.  For example, Fig.~\ref{fig:dkcr} illustrates that increasing $\hat b$ increases both $\hat d_{cr}$ and also $\hat k_{cr}$, although the qualitative behavior is unchanged.  In particular, $\hat k_{cr}$ still vanishes above a certain $n$.

We have seen that the $SL(2,{\mathbb{Z}})$-transformed theory with anyon statistics has a striped instability for ranges of $\hat d$ and $\hat b$ which are different from that for the original statistics.  However, as discussed in Sec.~\ref{sec:anyonization}, there is an alternate view of the $SL(2,{\mathbb{Z}})$ transformation in which the particle statistics remains the same but the dynamics of the fields change.  In the example of the S-transformed theory $n = \infty$, the charge current becomes fixed and the background gauge field becomes dynamical; this theory is interpreted as strongly-coupled $(2+1)$-dimensional QED. We focus now on this particularly interesting case.

 
\section{$2+1$ QED}
\label{sec:2+1QED}

After applying an $SL(2,{\mathbb{Z}})$ transformation to the original CFT, one can interpret the resulting theory as a new CFT with a conserved current carried by anyons.  The QNM spectrum we have computed corresponds to the poles of the two-point functions of this anyonic current. For the S-transformed theory, corresponding to $n \rightarrow \infty$,  there is another simple way of looking at things. The new theory is the original theory now coupled to a $U(1)$ vector field which is integrated in the path integral but has no Maxwell term.\footnote{One can generalize this to situations where the vector field has a Chern-Simons term.}

We can then interpret our model at $n \rightarrow \infty$ as describing QED, either for a large $N_f$ or at large coupling.  Of course, in addition to the $U(1)$ gauge field, the electrons have $SU(N)$ interactions as well. Let us see what we can learn about this theory from our computations.

If, in the initial theory, we start with a charge density $\hat{d}$ and magnetic field $\hat b = 0$, then after the S transformation, we will have a theory where the fermions are coupled to a dynamical gauge field with a fixed charge density $\hat{d}$. The QNMs computed in Sec.~\ref{sec:anyonicstripes} now correspond to the poles of the two-point correlation functions of the $U(1)$ field strength.  

As discussed in Sec.~\ref{sec:anyonicstripes}, for large $n$, the instability appears in the hydrodynamic diffusion mode. From the point of view of the gauge theory, the instability occurs in the two-point function of the physical photon field
\begin{equation}
\langle F_{xy} F_{xy}\rangle
\end{equation}
and therefore drives the theory towards a state of alternating stripes of positive and negative magnetic field. Note that the total magnetic field in all space is a conserved quantity in $2+1$ dimensions and so remains zero. As before, the nonlinear couplings cause other quantities become modulated as well.  In this case, the striped ground state also features a modulated transverse electric field $E_y$.

\begin{figure}[!ht]
\centering
\includegraphics[width=0.7\textwidth]{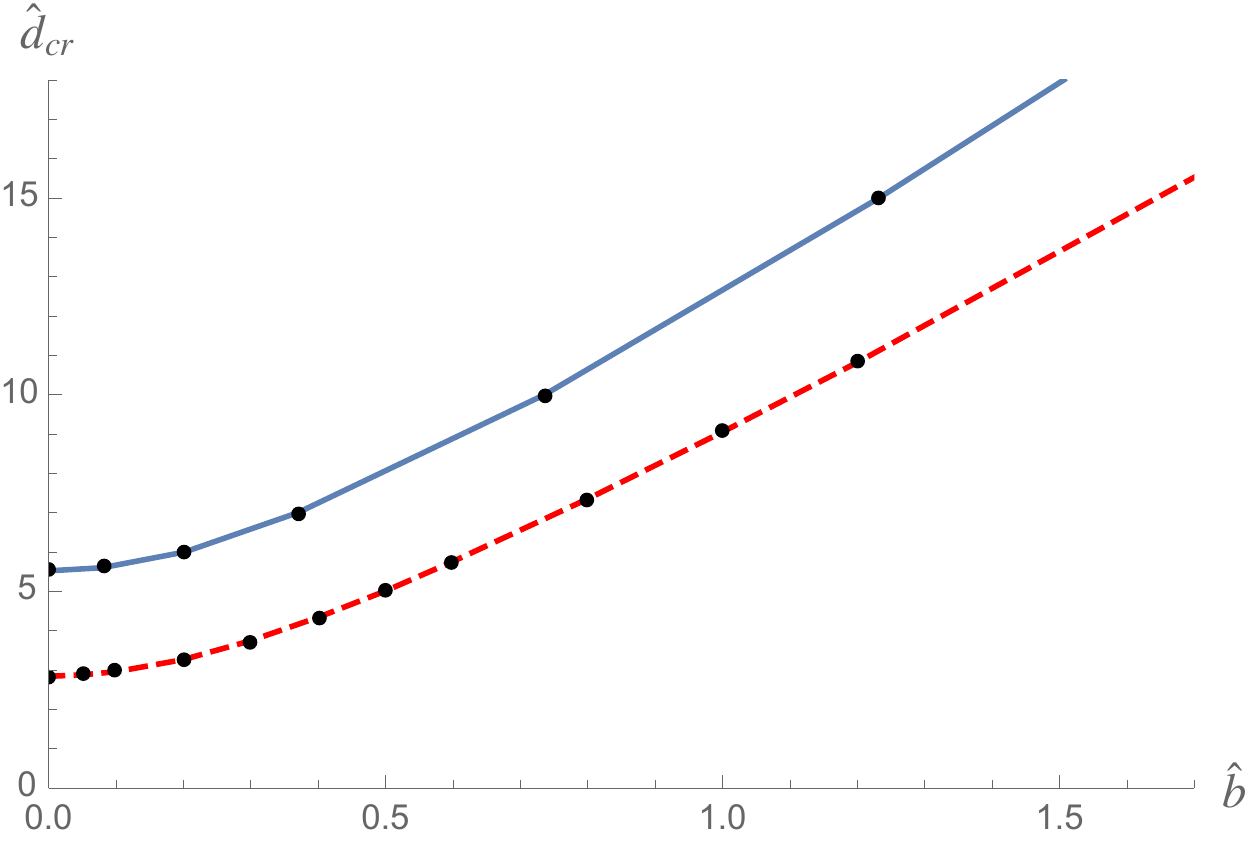}
\caption{The critical $\hat d_{cr}$ as a function of $\hat b$, for $n=0$ (solid, blue curve) and $n=\infty$ (dashed, red curve).  The curves are numerical fits to the black data ponts.}
\label{fig:dcrvsb} 
\end{figure}

As was seen in Fig.~\ref{fig:dkcr}, the $\hat d_{cr}$ at which the instability occurs goes down with increasing $n$.  This remains the case as $\hat b$ is increased above zero.  In Fig.~\ref{fig:dcrvsb} we show $\hat d_{cr}$ as a function of $\hat b$ for both $n=0$ and $n = \infty$.  As before, $\hat b$ acts to suppress the instability, and for a given $\hat{d}_{cr}$, there is a $\hat b$ above which the homogeneous phase is stable.

The fluctuation analysis of Sec.~\ref{sec:anyonicstripes} showed that $\hat k_{cr}$ vanishes above a certain $n$, as shown in Fig.~\ref{fig:dkcr}.  This suggests that the true ground state would feature a phase separation rather than a spatial modulation. However, this is not the case, as we can show by explicitly constructing the nonlinear modulated states, computing the free energy, and analyzing the phase diagram.

As discussed above, to compute the phase diagram for general $n$ as we did for $n=0$ in \cite{Jokela:2014dba} requires careful consideration of several boundary terms induced by the $SL(2,\mathbb{Z})$ mapping \cite{Ihl:2016sop}. While this is straightforward in principle, scanning through the whole parameter space is a daunting task requiring significant computational resources.
In this paper, we are content to focus on the case of $n=\infty$, which corresponds simply to imposing Neumann boundary conditions on the bulk gauge field.  In this case, $\hat d$ is held fixed, and it is as if we are just working in the canonical ensemble.

The standard quantization, $n=0$, corresponds to the grand canonical ensemble, with the chemical potential $\hat\mu = \hat a_0\big|_\partial$ kept fixed.  By the standard holographic dictionary, the grand canonical potential is just given by the Euclidean on-shell action with appropriate counterterms, $\Omega[\hat \mu] = S^{E}_{on-shell}$.  For $n=\infty$, the boundary conditions instead fix $\hat d \propto  \partial_r \hat a_0\big|_\partial$ (see equation \eqref{ddefintion}).  The free energy is  given by the Legendre transform of the grand canonical potential:
\be
F[\hat d] = S^{E}_{on-shell} + \hat d \hat \mu
\ee

To study the phase transition, we numerically solved the equations of motion for the modulated ground state, employing the techniques of \cite{Jokela:2014dba} but using the alternative boundary conditions. To determine the dynamically preferred spatial frequency $\hat k_0$ of the modulated state, we varied the fixed spatial periodicity $\hat L$ in order to minimize the free energy density $F/\hat L$.  We then computed the free energy density difference $\Delta F/\hat L$ between the modulated state and the homogeneous state.

\begin{figure}[!ht]
\centering
\includegraphics[width=0.45\textwidth]{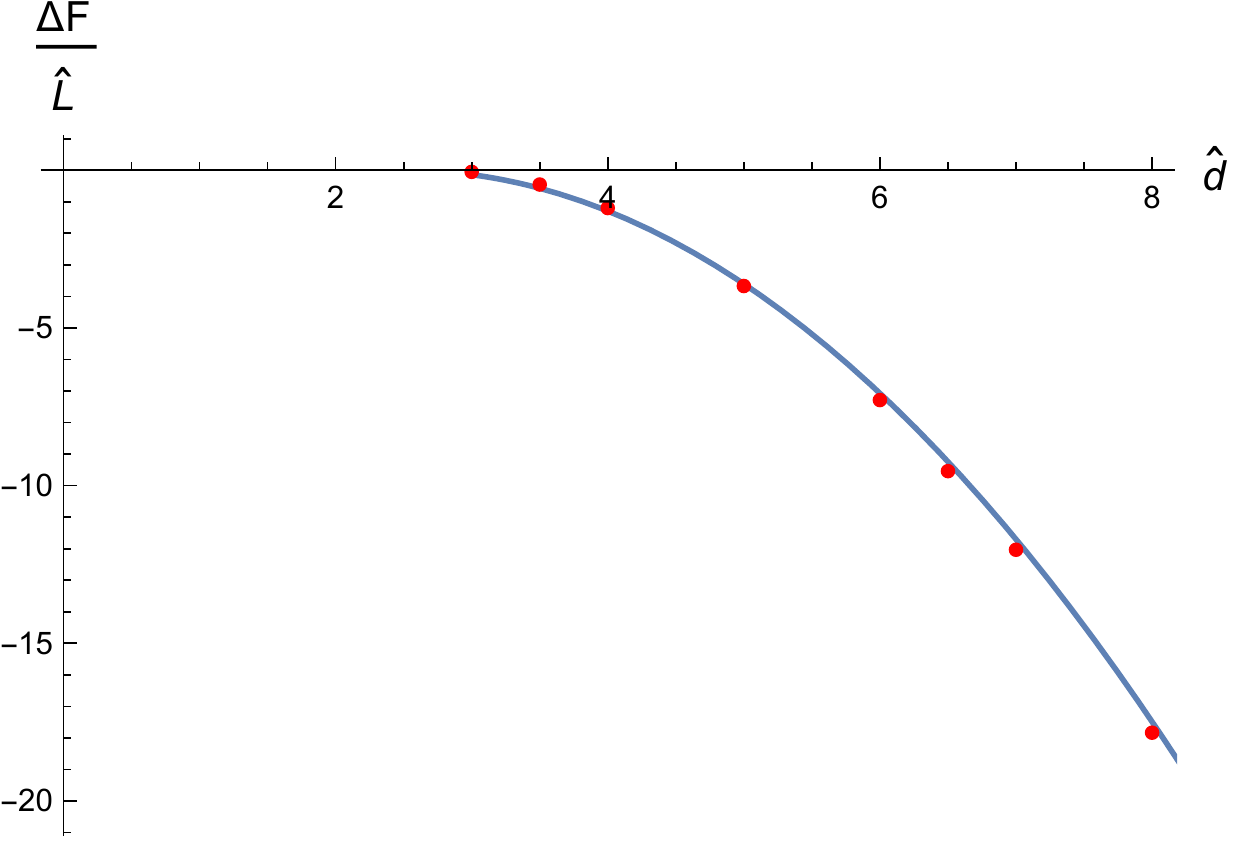}
\includegraphics[width=0.45\textwidth]{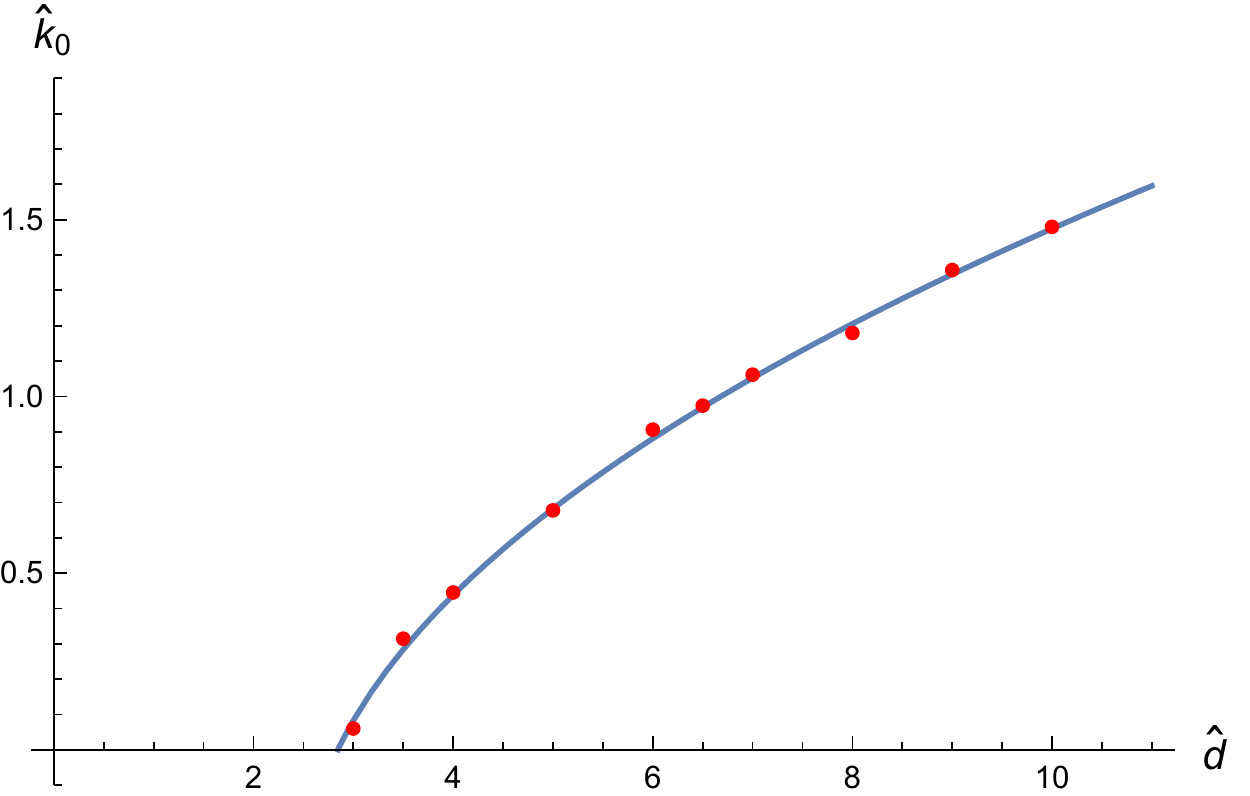}
\caption{(Left) The difference in free energy density $\Delta F$ between the homogeneous state and the modulated state plotted against $\hat d$ for $\hat b=0$.  The best-fit curve is $\Delta F/\hat L  = -0.6 (\hat d- \hat d_0)^2$, indicating a continuous phase transition at $\hat d_0 \approx 2.5$. (Right) The spatial frequency $\hat k_0$ of the modulated ground state plotted versus $\hat d$ in the S-transformed $n=\infty$ case. The best-fit curve is $\hat k_0 = 0.7 \sqrt{d-d_0-0.4}$.}
\label{fig:trueground} 
\end{figure}

The results for $\hat b = 0$ are shown in Fig.~\ref{fig:trueground}.  The left hand plot shows that, at small $\hat d$, $\Delta F/\hat L > 0 $, meaning the homogeneous phase is preferred. As $\hat d$ is increased, there is a transition at $\hat d = \hat d_{0} \approx 2.5$ to the modulated phase, as predicted by the perturbation analysis. The free energy density difference $\Delta F/\hat L $ smoothly approaches zero as $(\hat d- \hat d_{0})^2$, as shown in the left hand plot, implying a continuous phase transition. 

From the right hand plot, we see that for $\hat{d}$ well above $\hat{d}_{0}$, the true ground state has a nonzero $\hat k_0$, indicating stripes of finite wavelength. However, as $\hat{d}\rightarrow \hat{d}_{0}$, $\hat k_0$ trends toward zero.  Extrapolating the best-fit curve, we estimate $\hat{k}_0 \to 0$ at $\hat{d} \approx 2.8$ which is not quite equal to $\hat{d}_{0} \approx 2.5$, but within numerical error. It is numerically challenging to resolve the small $\hat k$ behavior, corresponding to large spatial periodicity $\hat L$.  Very large values of $\hat L$ are not well sampled by a finite number of grid points.  It is natural to conjecture that $\hat k_0 \rightarrow 0$ exactly at $\hat d_0$, which is certainly consistent with our results, given their numerical accuracy.  However, to verify this with a high degree of confidence will require more extensive numerical computations.

\section{Discussion and outlook}
\label{sec:discussion}

Anyons have the unique property that their statistics are tunable.  In this paper, we used holographic duality to study the effects of varying the statistics in a fluid of strongly interacting anyons, and we saw that varying the statistics parameter $n$ led to changes in the QNM spectrum and, in certain cases, in the qualitative characteristics of the ground state.

The effect of nonzero $n$ is in some ways similar to turning on a magnetic field, inducing a mixing of modes at small $k$ and leading eventually to a massive zero sound.  This is perhaps not surprising since turning on $n$ can be interpreted as turning on a statistical magnetic field.  However, increasing $n$ enhances, rather than suppresses, the modulated instability.  We also observed that the wavelength of the unstable mode grows with $n$, and for large enough $n$, the instability is no longer due to the transverse mode but to the hydrodynamical longitudinal mode.

The $SL(2,{\mathbb{Z}})$  transformation is typically understood as changing the particle statistics and mixing the charge current and background field of the boundary theory. 
However, as we saw, this procedure can be interpreted in an entirely different way; rather than acting on the particle statistics, the $SL(2,{\mathbb{Z}})$ transformation can be understood as altering the dynamics of the field theory.  

This view point is most applicable for the S-transformed theory, where $n=\infty$, in which the boundary gauge field becomes dynamical and the theory can be interpreted as strongly coupled, $(2+1)$-dimensional QED.  In this case, the instability is driven by fluctuations of the magnetic field and leads to a striped ground state in which the magnetic field is spatially modulated. We found that the phase transition at zero total magnetic field is still continuous but involves both the amplitude and the spatial frequency of the modulation vanishing at the critical point.

We conclude with several interesting open topics for future investigation. 
We found the conditions under which the homogeneous phase to become unstable and, for general $n$, inferred the presence of a nearby phase transition. However, we expect that, at least in certain regions of parameter space, that the true ground state is neither homogeneous nor striped but rather a checkerboard, with continuous translation symmetry broken in both directions, as was found in \cite{Withers:2014sja, Donos:2015eew}.  As a first step toward inferring the existence of such a phase, one could attempt to study fluctuations of the striped states, looking for further symmetry-breaking instabilities. 
 
For the case of $n=\infty$, we showed in Sec.~\ref{sec:2+1QED} that the correct free energy was found by adding a boundary term corresponding to a Legendre transformation.  In general, an $n$-dependent boundary term must be added \cite{Ihl:2016sop}, and the resulting free energy for the ground state can be obtained. We leave this generalization for future work.

Electrical conductivity is an important observable that provides a useful window on the behavior of the anyonic fluid.  The DC conductivity of original, homogeneous D3-D7' model with $n=0$, was computed in \cite{Bergman:2010gm} by the Karch-O'Bannon method, and the DC and AC conductivities of both the homogeneous phase and the striped phase were computed in \cite{Jokela:2016xuy}.  The conductivity for general $n$ could be computed by the straightforward boundary condition modification described in Sec.~\ref{sec:fluctuations}.  This result would be expected to match the general formula conjectured in \cite{Ihl:2016sop}, which follows from requiring the corresponding Green's functions to transform covariantly under $SL(2,\mathbb{Z})$.
 
In this paper, we have focused on the gapless, conducting phase given by the black hole embeddings of the D3-D7' model.  The Minkowski embeddings were the subject of two earlier papers \cite{Jokela:2013hta, Jokela:2014wsa}, which found that for a particular choice of $n$, the gapped quantum Hall state was transformed into an anyonic superfluid.  It would be interesting to understand how the system transitions from the metallic phase studied here to the superfluid.\footnote{The choice of internal fluxes made here, $f_1 = f_2 = \frac{1}{\sqrt 2}$, does not admit Minkowski embeddings.  In order to study the transition from black hole to Minkowski embedding, either $f_1$ or $f_2$ would have to be set to zero.} 

Finally, one long-standing question in holography is how to directly observe the statistics of the strongly interacting particles of the boundary theory.  In a perturbative description, single-particle operators can be used to probe the properties of the underlying particles. The holographic description, instead, is in terms of collective excitations, like energy and charge currents, and a quasiparticle description is lacking.  Of course, this is not at all surprising, given that particles with tractable holographic duals are strongly interacting.  
 
In a top-down holographic model, one can infer the particle statistics from the weakly coupled, flat-space brane intersection description.  In the D3-D7' model, for example, the lowest excitations of the 3-7 open strings are fundamental fermions, which become the particles of the boundary field theory in the decoupling, near-horizon limit.  But, the question remains: how do the bulk observables encode the statistics of the boundary particles?


\vspace{0.8cm}

{\bf \large Acknowledgments}
We thank Matti J\"arvinen, Esko Keski-Vakkuri, and Tobias Zingg for discussions.
N.J. is supported by the Academy of Finland grant no. 1297472.
M.L. was supported by funding from the European Research
Council under the European Union's Seventh Framework Programme (FP7/2007-2013) /
ERC Grant agreement no.~268088-EMERGRAV.  
M.L. was also supported by a professional development grant from Long Island University.
The work of GL is supported in part by the Israel Science Foundation under grant 504/13.
We thank the Galileo Galilei Institute for Theoretical Physics for the hospitality and the INFN for partial support while this work was in progress.


\appendix

\section{Background equations of motion}
\label{app:backgroundeom}

In this appendix, we reproduce the equations of motion for the D7-brane embedding scalars $\psi(r)$ and $z(r)$ and radial worldvolume gauge field $a_t(r)$, which were originally derived in \cite{Bergman:2010gm}.

The action has no explicit dependence on $z$, so the equation of motion for $z(r)$ can be integrated once to give
\be
\label{zeqn}
 z' = \frac{\frac{c_z}{r^4}-f_1 f_2}{h}
 \sqrt{ \frac{1+hr^2\psi'^2}{\tilde{d}^2+\frac{r^4}{h}\left(h\left(1+\frac{b^2}{r^4}\right)\left(f_1^2+4\cos^4\psi\right)\left(f_2^2+4\sin^4\psi\right)-\left(\frac{c_z}{r^4}-f_1 f_2\right)^2\right) }   }  \ ,
\ee
where $c_z$ is a constant of integration fixed by regularity at the IR boundary.  For black hole embeddings, $c_z = f_1 f_2 r_T^4$.

The equation of motion for $a_t(r)$ can also be integrated once to give
\be\label{Aeqn}
 a'_t = \tilde{d}\sqrt{ \frac{1+hr^2\psi'^2}{\tilde{d}^2+\frac{r^4}{h}\left(h\left(1+\frac{b^2}{r^4}\right)
 \left(f_1^2+4\cos^4\psi\right)\left(f_2^2+4\sin^4\psi\right)  -\left(\frac{c_z}{r^4}-f_1 f_2\right)^2\right)  } } \ ,
\ee
where $\tilde d(r) \equiv d - 2bc(r)$ is the radial electric displacement, the constant of integration $d$ is the boundary charge density, and the function $c(r)$ is defined in equation \eqref{cr}.

Finally, the equation of motion for $\psi(r)$ is
\bea
\label{psieom3}
 & & \partial_r\left( r^2 g(r)\left(1 + {b^2\over r^4}\right) \psi'(r)  \right) = -16b\cos^2\psi\sin^2\psi \  a'_t
 \nonumber \\
       &&   \mbox{}  + \frac{8hr^4}{g(r)} \cos\psi\sin\psi \left[(f_1^2 + 4\cos^4\psi)\sin^2\psi 
           - (f_2^2 + 4\sin^4\psi)\cos^2\psi\right] \ ,
\eea
where the function $g(r)$ is given by
\be 
 g = {h\over \left(1+{b^2\over r^4}\right)}
  \sqrt{\frac{\tilde{d}^2+\frac{r^4}{h}\left(h\left(1+\frac{b^2}{r^4}\right)\left(f_1^2+4\cos^4\psi\right)\left(f_2^2+4\sin^4\psi\right)-\left(\frac{c_z}{r^4}-f_1 f_2\right)^2\right)}{1+hr^2\psi'^2}}  \ .
\ee

\section{Fluctuation equations of motion}
\label{app:fluctuationeom}

This appendix contains the equations of motion for the fluctuation fields $\delta\tilde \psi$, $\delta \tilde z$, and $\delta \tilde a_\mu$. Despite being rather long, these equations are straightforward to obtain and were previously worked out in \cite{Jokela:2012vn}.

First, let us define the functions
\bea
 \hat g  & \equiv & \frac{h}{1+\hat b^2 u^4}\sqrt{\frac{\hat{\tilde d}^2 u^4+(1+\hat b^2 u^4)G-h f_1^2 f_2^2}{1+h u^2\psi'^2}} \\
 G & \equiv & \left(f_1^2+4\cos^4\psi\right)\left(f_2^2+4\sin^4\psi\right) \\
 A & \equiv & 1+h u^2\psi'^2+hu^{-4}\bar z'^2-\bar a'^2_0 \ ,
\eea
where $\hat{\tilde d}\equiv \hat d-2c(u)\hat b$ and the prime denotes differentiation with respect to $u$.  However, for the background fields $z$ and $a_0$, we have also defined $\bar a_0 '\equiv \partial_r a_0 =  - \frac{u^2}{r_T} \partial_u a_0$ and $\bar z' \equiv \partial_r z =  - u^2 r_T  \partial_u z$.
We also define
\bea
  \hat H & \equiv &  \frac{\hat g u^2}{A h}\left(1+\hat b^2 u^4\right)\left(1+hu^{-4}\bar z'^2+hu^2\psi'^2\right)\delta \hat a'_t \nonumber \\
 & & -\left(4\hat b\sin^2(2\psi)+\frac{\hat g}{2h}\bar a'_0\left(1+\hat b^2 u^4\right)\partial_\psi\log G\right)\delta\tilde\psi\nonumber\\
 & & +\frac{\hat gu^2}{A}\left(1+\hat b^2 u^4\right)\bar a'_0\psi'\delta\tilde\psi' -\frac{\hat g}{Au^2}\left(1+\hat b^2 u^4\right)\bar a'_0 \bar z'\delta\tilde z' \ .
\eea

The $\delta\tilde\psi$ equation of motion reads:
\bea
 & & \left(-\frac{h}{2\hat gu^4}\left(\partial_\psi^2 G-\frac{1}{2G}(\partial_\psi G)^2\right)+8\hat b\bar a'_0\sin(4\psi)+\frac{u^2}{2}\partial_u\left(\hat g\psi'\left(1+\hat b^2 u^4\right)\partial_\psi\log G\right)   \right)\delta\tilde\psi \nonumber\\
 & & = -u^2\partial_u\left(\frac{\hat g}{A}\left(1+\hat b^2 u^4\right)\left(1+hu^{-4}\bar z'^2-\bar a'^2_0\right)\delta\tilde\psi'\right)\nonumber\\
 & & +\frac{\hat gu^2}{h^2}\left(-\left(1+\hat b^2 u^4\right)\left(1+hu^{-4}\bar z'^2\right)\hat\omega^2+\left(1+hu^{-4}\bar z'^2-\bar a'^2_0\right)h\hat k^2   \right)\delta\tilde\psi\nonumber\\
 & & -\frac{\hat g}{2u^2}\left(1+\hat b^2 u^4\right)\partial_\psi\log G\bar z'\delta\tilde z'+\frac{\hat g}{h}\bar z'\psi'\left(-\left(1+\hat b^2 u^4\right)\hat\omega^2+h\hat k^2\right)\delta\tilde z\nonumber\\
 & & -u^2\partial_u\left(\frac{\hat g h}{Au^2}\left(1+\hat b^2 u^4\right)\bar z'\psi'\delta\hat z'\right)+\left( 4\hat b\sin^2(2\psi)+\frac{\hat g}{2h}\bar a'_0\left(1+\hat b^2 u^4\right)\partial_\psi\log G   \right)u^2\delta \tilde a'_0 \nonumber\\
 & & +u^2\partial_u\left(\frac{\hat gu^2}{A}\bar a'_0\psi'\left(1+\hat b^2 u^4\right)\delta\tilde a'_0  \right)-\frac{\hat gu^4}{h}\bar a'_0\psi'\hat k\delta \tilde e_x\nonumber\\
 & & -i\hat k\left( 4\bar a'_0\sin^2(2\psi)+\frac{h\hat b}{2\hat g\left(1+\hat b^2 u^4\right)}\partial_\psi G-\hat b u^2\partial_u\left(gu^4\psi' \right)  \right)\delta\tilde a_y \ .
\label{deltapsiEOM}
\eea

The $\delta\tilde z$ equation of motion reads:
\bea
 & & 0 = \frac{\hat g}{h}\bar z'\psi'\left(-\left(1+\hat b^2 u^4\right)\hat\omega^2+h\hat k^2\right)\delta\tilde\psi \\
 & & -u^2\partial_u\left(\frac{\hat g h}{Au^2}\left(1+\hat b^2 u^4\right)\bar z'\psi'\delta\tilde\psi'-\frac{\hat g}{2u^4}\partial_\psi\log G\bar z'\delta\tilde\psi\right) \nonumber\\
 &  & -u^2\partial_u\left(\left(1+\hat b^2 u^4\right)\frac{1+hu^2\psi'^2-\bar a'^2_0}{A u^2}\hat g\delta\hat z'\right)\nonumber\\
 & & +\frac{\hat g}{h^2}\left(-\left(1+\hat b^2 u^4\right)\left(1+hu^2\psi'^2\right)\hat\omega^2+\left(1+hu^2\psi'^2-\bar a'^2_0\right)h\hat k^2   \right)\delta\tilde z \nonumber\\
 & & +\frac{\hat g}{h}\bar a'_0\bar z'\hat k\delta\tilde e_x-u^2\partial_u\left(\frac{\hat g}{Au^2}\left(1+\hat b^2 u^4\right)\bar a'_0\bar z'\delta\tilde a'_0\right)\nonumber\\
 & & -i\hat k\hat b u^2\delta\tilde a_y\partial_u\left(\hat g\bar z'\right) \ .
 \label{deltazEOM}
\eea

The $\delta\tilde a_0$ equation of motion reads:
\bea
 & & 0 = u^2  H-\frac{\hat g}{h}u^4\bar a'_0\psi'\hat k^2\delta\tilde \psi+\frac{\hat g}{h}\bar a'_0\bar z'\hat k^2\delta\tilde z \nonumber\\
 & & -\hat k\frac{\hat g}{h^2}u^4\left(1+hu^{-4}\bar z'^2+hu^2\psi'^2\right)\delta\tilde e_x-i\hat k\delta\tilde a_y u^2\partial_u\left(2 c(u)-\frac{\hat b\hat g}{h}u^4\bar a'_0\right) \ .
\label{deltaatEOM}
\eea

The $\delta \tilde a_x$ equation of motion reads:
\bea
 & & 0 = -\frac{\hat g}{h}u^4\psi'\bar a'_0\hat k\hat\omega\delta\tilde\psi+\hat k\hat\omega\frac{\hat g}{h}\bar a'_0\bar z'\delta\tilde z-\hat\omega\frac{\hat g}{h^2}u^4\left(1+hu^{-4}\bar z'^2+hu^2\psi'^2\right)\delta\tilde e_x\nonumber\\
 & & -i\hat\omega\delta\tilde a_y u^2\partial_u\left(2c(u)-\frac{\hat b\hat g}{h}u^4\bar a'_0\right)+\frac{u^2}{\hat\omega}\partial_u\left(\hat g u^2\left(-\delta\tilde e'_x+\hat k\delta\tilde a'_t\right)\right) \ .
\label{deltaaxEOM}
\eea

The $\delta\tilde a_y$ equation of motion reads:
\bea
 & & 0 = i\hat k\hat b\delta\tilde\psi u^2\partial_u\left(\hat g u^4\psi'\right)+4i\hat k\bar a'_0\sin^2(2\psi)\delta\tilde\psi+i\hat k\frac{h\hat b\partial_\psi G}{2\hat g\left(1+\hat b^2 u^4\right)}\delta\tilde\psi \nonumber\\
 &  & +i\hat k\hat b\delta\tilde z u^2\partial_u\left(\hat g\bar z'\right)+i\delta\tilde e_x u^2\partial_u\left(2c(u)-\frac{\hat b\hat g}{h}u^4\bar a'_0\right)-u^2\partial_u\left(\hat g u^2\delta\tilde a'_y\right)\nonumber\\
 & & -\frac{\hat g}{h^2}u^4\left(1+hu^{-4}\bar z'^2+hu^2\psi'^2\right)\hat\omega^2\delta\tilde a_y+\frac{\hat g}{h}u^4 A \hat k^2\delta \tilde a_y \ .
\label{deltaayEOM}
\eea

And finally, the constraint coming from the $\delta a_u$ equation of motion, which maintains the gauge condition $a_u=0$, reads:
\be
 -\hat\omega H+\frac{\hat k}{\hat\omega}u^2\hat g\left(-\delta \tilde e'_x+\hat k\delta\tilde a'_0\right) = 0 \ .
\label{deltaauEOM}
\ee


\end{document}